# Machine learning and high-throughput robust design of P3HT-CNT composite thin films for high electrical conductivity

Daniil Bash*[1,2], Cai Yongqiang*[3,4], Vijila Chellappan*[1], Wong Swee Liang[1], Xu Yang[1], Pawan Kumar[1], Tan Jin Da[1,5], Anas Abutaha[1,6], Jayce Cheng[1], Lim Yee Fun[1], Siyu Tian[7], Danny Zekun Ren[7], Flore Mekki-Barrada[8], Wai Kuan Wong[8], Jatin Kumar[1], Saif Khan[8], Li Qianxiao[#3,9], Tonio Buonassisi[#7,10], Kedar Hippalgaonkar[#1,11]

*equal contribution, correspondence to qianxiao@nus.edu.sg , buonassisi@mit.edu, kedar@ntu.edu.sg

[1] Institute of Materials Research & Engineering, Agency for Science Technology and Research, 2 Fusionopolis Way, 138634 Singapore
[2] Department of Chemistry, 3 Science Drive 3, National University of Singapore, 117543 Singapore
[3] Department of Mathematics, National University of Singapore, 21 Lower Kent Ridge Rd, 119077 Singapore
[4] School of Mathematical Sciences, Laboratory of Mathematics and Complex Systems, MOE, Beijing Normal University, 100875 Beijing, China
[5] NUS Graduate School for Integrative Sciences & Engineering (NGS), University Hall, Tan Chin Tuan Wing Level 04, #04-02, 21 Lower Kent Ridge Road, 119077, Singapore
[6] Qatar Environment and Energy Research Institute, Hamad Bin Khalifa University, Qatar Foundation, Doha, 34110, Qatar
[7] Singapore-MIT Alliance for Research and Technology SMART, 138602, Singapore
[8] Department of Chemical and Biomolecular Engineering, National University of Singapore, 117585, Singapore.
[9] Institute of High Performance Computing, 1 Fusionopolis Way, #16-16, Connexis 138632, Singapore
[10] Massachusetts Institute of Technology, Cambridge, MA 02139, USA.
[11] Department of Materials Science and Engineering, Nanyang Technological University, Singapore 639798, Singapore

# Abstract

Combining high-throughput experiments with machine learning allows quick optimization of parameter spaces towards achieving target properties. In this study, we demonstrate that machine learning, combined with multi-labeled datasets, can additionally be used for scientific understanding and hypothesis testing. We introduce an automated flow system with high-throughput drop-casting for thin film preparation, followed by fast characterization of optical and electrical properties, with the capability to complete one cycle of learning of fully labeled ~160 samples in a single day. We combine regio-regular poly-3-hexylthiophene with various carbon nanotubes to achieve electrical conductivities as high as 1200 S/cm. Interestingly, a non-intuitive local optimum emerges when 10% of double-walled carbon nanotubes are added with long single wall carbon nanotubes, where the conductivity is seen to be as high as 700 S/cm, which we subsequently explain with high fidelity optical characterization. Employing dataset resampling strategies and graph-based regressions allows us to account for experimental cost and uncertainty estimation of correlated multi-outputs, and supports the proving of the hypothesis linking charge delocalization to electrical conductivity. We therefore present a robust machine-learning driven high-throughput experimental scheme that can be applied to optimize and understand properties of composites, or hybrid organic-inorganic materials.

# 1 Introduction

One of the biggest bottlenecks for fast and efficient scientific discoveries is the bandwidth and cognitive ability of human researchers, who can conceive, conduct and comprehend a limited number of experiments. With the emergence of automation, timelines for scientific progress can be significantly accelerated via the implementation of high-throughput experiments. From the first automated pipette in 1950[1] to modern self-driving laboratories[2–4], automation tools allow researchers to explore complex, multi-dimensional parameter spaces, while freeing up researchers' bandwidth for planning experiments and analyzing data. Coupled with the emergence of machine learning as a cognitive assistant, to plan and navigate complex parameter spaces, a new paradigm has emerged that is especially useful for combinatorial experiments[5–9] and multi-variable optimization problems[10–12]. These methods have been used for discovery of novel metal alloys[9,10,13–15]. perovskite materials for photovoltaics[2,11,16], electronic property optimization of polymer thin films[2,3], or synthesis of co-polymers with pre-defined target properties[17].

In this report, we present a high throughput, semi-automated experimental platform driven by machine learning to maximize the electrical conductivity of inorganic-organic hybrid materials. As an integral part of this platform, the machine learning algorithm first performs Bayesian Optimization (BO) for targeted sampling of data, which are then used to build and select robust graphical neural network models that link multiple experimental inputs to outputs. These can then be used for analysis and design of high-performance materials for desired applications.

This machine learning driven high throughput experimental platform is used to demonstrate optimization of electrical conductivity in regio-regular poly-3-hexyl thiophene (rr-P3HT) and carbon nanotube (CNT) composites. The composite films of conjugated polymers with single or multiwall carbon nanotubes (SWCNTs, MWCNTs) have been used as an active material in various functional devices due to their unique optical, electrical and mechanical properties[18–21]. In particular, the conjugated polymer/CNT composites combine the advantages of mechanical flexibility and low-cost manufacturing processes making them suitable for electronic and thermoelectric applications. P3HT/CNT is one such promising nano-composite material but the electrical conductivity ($\sigma$) requires significant improvement in order to make this class of nanocomposites viable for practical applications[22]. Various strategies are proposed to increase the electrical conductivity of P3HT/CNT composites including molecular doping of P3HT, using different CNT types[23], optimizing process, tuning the energy barrier between the polymer and CNT[24] and improving the crystalline structure or morphology of P3HT and CNT[25–27]. Most of these approaches are concentrated on enhancing inter-chain charge transport by improving the degree of crystallinity or nanoscale architecture of P3HT chains that are induced by CNTs[28]. The configuration of P3HT wrapping around the CNTs, particularly the formation of elongated polymer chain conformation with reduced torsional disorder, promotes inter-chain ($\pi$-$\pi$ stacking) interaction and thus increase charge mobility[19]. However, the degree of crystallinity and orientation of $\pi$-$\pi$ stacking depends on the interaction between P3HT and CNT where the parameters such as type, size and length of CNT manipulates the crystalline packing structure thus influencing the charge transport[19,23]. In addition to the CNT-induced charge carrier mobility enhancement, the chemical doping of P3HT increases the electrical conductivity since doping of conjugated polymer with small molecular oxidizers such as tetra cyano quinodimethane (F4TCNQ) or iron(III) chloride ($FeCl_3$) introduces mobile charges to the conjugated polymer chain that increases charge carrier concentration and reduces π-π stacking distance due to structural reorganization and polaron delocalization[29–31]. Therefore, there is a rich interaction space available by changing the type of CNTs and tuning the physical and chemical interactions to enhance the electrical conductivity.

In this study, we mix rr-P3HT with four types of CNTs, where the interaction between the P3HT polymer chains and CNTs are expected to create different composite morphologies and optimal crystalline structure for maximizing electrical conductivity. The types of CNTs used in this study are: 1. Long single wall CNT of lengths in the range of 5 - 30 μm (*l*-SWNTs), 2. Short single wall CNT of lengths in the range of 1 - 3μm, (*s*-SWNTs), 3. Multi wall CNT (MWCNTs) and 4. Double-wall CNT (DWCNTs). The details of the materials can be found in supplementary information (SI), Section 1.1. All the composite films are doped with $FeCl_3$ in order to increase the overall electrical conductivity and it has already been reported that the dopant is insensitive to the types of CNT as doping increases the electrical conductivity of all the P3HT/CNT composites irrespective of the type of CNT[23]. The workflow begins with data generation from the high-throughput experimental platform where P3HT/CNT composite films are prepared by a microfluidic reactor combined with an automated drop-casting system, fast optical and electrical diagnostics, and the use of Bayesian Optimization to scan the experimental data manifold, explore input-output correlations and target high conductivity.

## 2 Data Generation: High Throughput Experiments with BO

We first apply Bayesian Optimization (BO)[32,33] to optimize high-throughput processing and characterization and explore the parameter space for P3HT-CNT composites. This work is based on a hypothesis that a certain combination of CNTs and P3HT should sensitively affect the alignment of P3HT chains to change the conductivity of the system, while dispersed nanotubes, due to their unique dimensionality would act as seeds for P3HT crystallization. This hypothesis is qualitatively explored via high throughput experimentation and machine learning. The schematic representation of the entire workflow is shown in figure 1 that involves data generation from the high-throughput experimental platform, (a) film fabrication, (b) fast optical/electrical labeling, (c) use of Bayesian Optimization to for targeted exploration of the composition space, and construction of ML models for (d) predicting electrical conductivity through graph based models and (e) to derive correlation between the composition and electrical conductivity for interpretable ML. The high throughput experimental platform consists of a labview-controlled automatic flow processing system where formulations of different starting materials with varying composition ratios and concentrations are mixed *in situ*. The mixing volume and ratio between the pristine, stable stock solutions (details on how these solutions are prepared can be found in *Supplementary Information, Section 1.2*) are adjusted in order to prepare unique formulations. Each formulation forms a droplet and multiple droplets are subsequently drop-casted onto a pre-cleaned smooth, double-side polished fused silica (quartz) wafer, eliminating the time-consuming process of cleaning, surface treating individual substrates and film preparation. The optical and electrical properties of the droplets are measured using a high-throughput diagnostic platform consisting of a visible range (400-1000nm) hyperspectral imaging system and a computer-controlled automated four-point probe setup (Figure 1b). This platform is driven by a BO algorithm for optimising the electrical and optical characteristics and related to the input processing/mixing conditions.

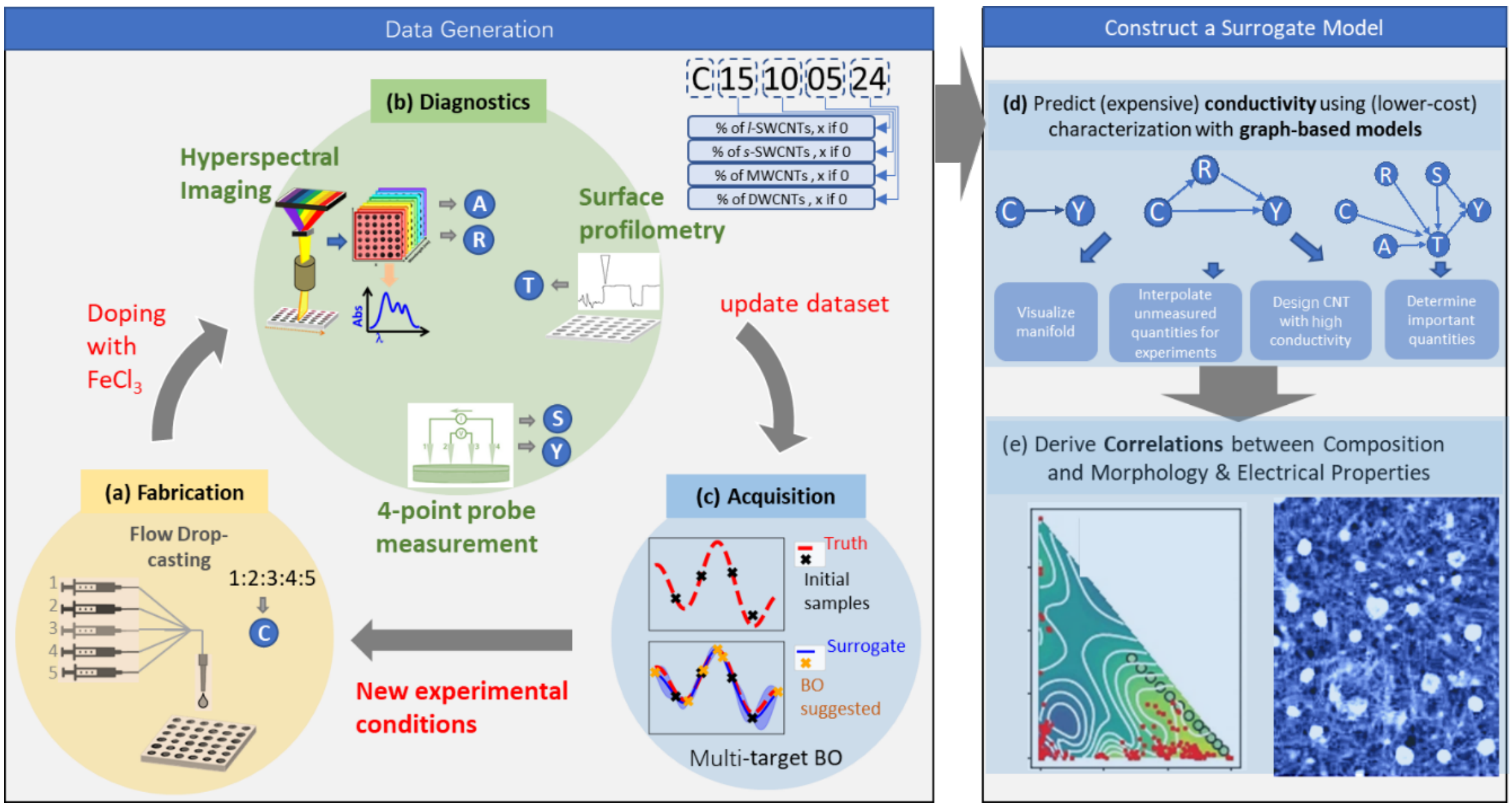

*Figure. 1 Schematic representation of the workflow involving high-throughput experimental platform for film data generation (a-c) and machine learning based data analysis (d-e).* ***(a)***. Film fabrication. *The workflow begins with the preparation of hybrid solution by mixing the pre-prepared P3HT and P3HT/CNT stock solutions (1mg/ml in o-Dichlorobenzene) using an automated microfluidic flow reactor. The micro fluidic flow reactor is capable of mixing 5 different liquids in a precise ratio and separates them as 'plugs' containing unique composition ratio (C) of P3HT and CNT. Five liquids include pure P3HT solution, and four P3HT/CNT mixtures containing different types of CNTs. The unique P3HT/CNT mixtures are then drop casted on pre-treated quartz wafer placed on a computer controlled XY stage. 36 droplets that contain 6 unique compositions are drop casted on a single quartz wafer where each condition has 6 replicates in order to check the reproducibility of the droplets. The drop-casted films are doped with $FeCl_3$ after recording the Hyperspectral image of pristine films.* ***(b)*** fast optical/electrical diagnostics: *Hyperspectral image (Spectral region from 400 to 1000nm) of the drop casted pristine films (before doping) and doped films are obtained. The absorption spectra of every droplet are obtained by averaging ~20 pixels (area of 2.4 mm) in the middle of the droplet. The absorbance ratio (R) between the interchain transition due to $\pi$-$\pi$ stacking (~602 nm) and intrachain transition due to $\pi$-$\pi^*$ are recorded as one of the proxy measurements to relate to the electrical conductivity of the film. The absorption spectra of the doped films are also recorded where the absorbance at 525nm (A) is recorded for evaluating the thickness of the doped film. The sheet resistance (S) of the doped films is tested using automated four-point probe measurement and then the doped films undergone Surface Profilometer where the film thickness (T) of 15 % of the droplets is measured. The electrical conductivity ($\sigma$) of the doped films are estimated using the measured/evaluated thickness value from the Profilometer and absorbance (A);* ***(c)****. Use of Bayesian optimisation (BO) to explore the electrical and optical characteristics related to the input processing/mixing conditions. The surrogate model in BO is a Gaussian process regressor. (d) Construction of graph-based models taking the absorbance, absorbance ratio, and sheet resistance into account to exploit maximal useful information when predicting electrical conductivity. (e) The selected graphs are appilied to derive correlation between the composition and electrical conductivity for interpretable machine learning.*

## 2.1 High Throughput Experiments

The hybrid solutions are mixed using a microfluidic flow reactor in order to fully exploit the composition space (C). The reactor consisted of 6 computer-controlled syringe pumps, mixing chamber, gas-inlet T-junction, and gas flow controller. This setup allowed to mix 5 different stock solutions together in precise ratios and separate the flow into "plugs" using $N_2$ gas. It uses a microfluidic flow reactor to mix the precursor solutions, and then drop-cast them onto the substrates in an automated high-throughput fashion, via a computer-controlled XY stage. The samples then undergo post–treatment in the form of doping, to enhance the electrical conductivity of the polymer matrix (Refer SI, section 1.4 for more details).

The drop casted films before and after doping are screened using Hyperspectral imaging (HSI) system that measures the optical absorbance spectra in the visible to near infrared region (400-1000 nm). The detailed HSI measurement protocols can be found in SI, section 2.1.The absorption spectra of P3HT/CNT composite films without post treatment are used to probe the degree of π-π stacking of P3HT. The absorption spectrum of P3HT/CNT composite films generally shows a peak at 525 nm, which corresponds to the intra-chain π–π* transition and the peaks at 550 nm and 602 nm correspond to vibronic and interchain transition due to π-π stacking respectively (refer SI, figure S1 & S2 for more details). The intensity of inter-chain transitions increases when the number of well-ordered P3HT aggregates increases. Therefore, we use absorption ratio (R) (ratio between the inter-chain interaction and intra-chain interaction) as one of the labeled measurements to correlate to the electrical conductivity ($\sigma$). The absorption ratio (R) of pristine films along with composition ratio (C) is used to train the ML algorithm in order to increase the degree of π-π stacking, which is hypothesized to increase the electrical conductivity of the P3HT/CNT composite. The hyperspectral images of the doped film are also measured in order to obtain the absorbance of the doped film where the absorbance (A) at 525nm is used to estimate the thickness of the doped films . Then the sheet resistances (S) of the doped films are measured using automated 4-point probe measurements, following which the films undergo surface profilometry measurement where the thickness (T) of 15 % of the droplets are measured. The electrical conductivity ($\sigma$) of all the droplets are obtained using the measured and estimated film thickness (T). Thus, our experimental workflow can rapidly measure the structural, optical and electrical labels of prepared samples, which are then used to create the dataset for machine learning.

## 2.2 Data Generation Guided by Batch Bayesian Optimization

The dataset $\mathcal{D}$ consists of experimental volumes of each CNT type and P3HT, that make up the film, as well as its associated properties measured. These properties include optical absorbance (A), absorption ratio ([R]), sheet resistance ([S]), film thickness ([T]), and film conductivity ($\sigma$), as shown in Table I. Before using the machine learning method, we need a pretreatment for those parameters, particularly, we take logarithms on [R],[S],[T], $\sigma$ and denote them by 'R','S','T','Y', which makes the measurement noise closer to additive independently, identically distributed (i.i.d.) noise (see SI figure S3). To account for variance, we drop-casted six films with identical experimental inputs (C) and measured the properties of the generated films.

| Attribute | Symbol |
|---|---|
| Composition | $C$ |
| Absorbance | A |
| Ln(Abs. Ratio) | R |
| Ln(Sheet Res.) | S |
| Ln(Thickness) | T |
| Conductivity | $\sigma$ |
| Log conductivity | $Y$ |

*Table I. Attributes and symbols used for the attributes for multi-labelled data generation for inputs (C) and outputs (A, R, S, T, $\sigma$ and Y)*

Next, we build our dataset through a combined batch Bayesian Optimization (BO) and regression model, as shown in Figure 2. BO is a heuristic global optimization algorithm that incorporates exploration in the parameter space. Here, we use BO, by GPyOpt package[34],mainly for a guided exploration of parameter regions that correspond to both high and low values of R and Y, to build a dataset representative of the whole landscape.

Due to expensive measurement costs associated with T and Y, we only measure around 15% (randomly chosen) of them. The rest of the properties (A, R, S) can be obtained in high-throughput hence we measure all of them. We then use a regression model to populate the unmeasured values in Y, after which the BO step can continue. After every step, the regression model is updated with the newly measured values, thereby building the dataset and reducing the uncertainty. Initially, when the dataset is small, a simple linear regression model (C→Y) is used to ensure good generalization. As more data is collected, we transition to a gradient boosting model to increase accuracy and reduce bias.

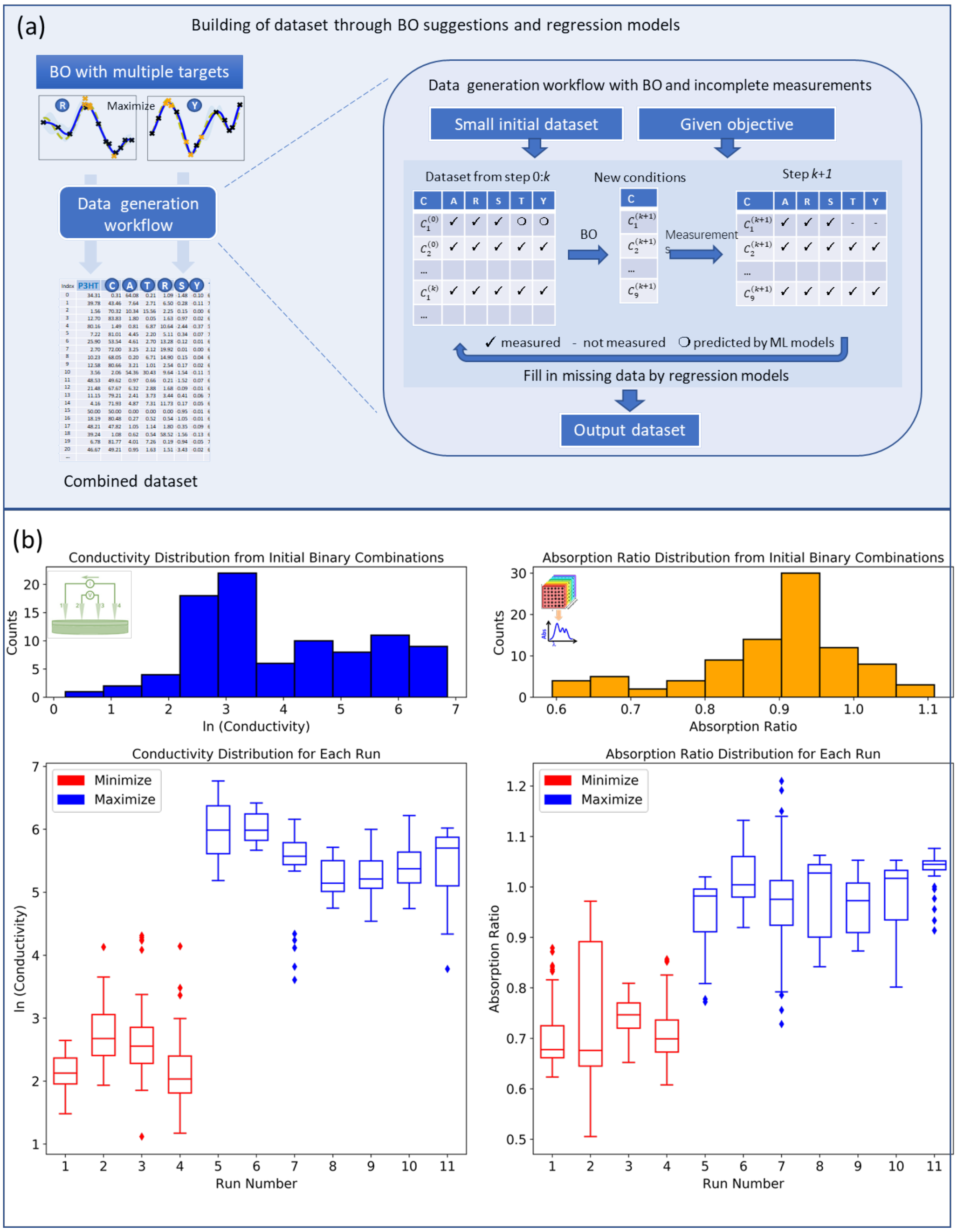

*Figure 2: Building of dataset through batch BO and regression models. The initial dataset is small and has data corresponding to composition $C_1^{(k)}, C_2^{(k)}$, etc. $(k = 0)$. The dataset is updated by the 'data generation workflow' in (a) with multiple targets (R and Y, min and max).*

*For a given target, batch BO (GPyOpt package, batch size=9) will suggest new conditions $C_1^{(k+1)}, \dots, C_9^{(k+1)}$, which will be added in the updated dataset after measuring the properties. Each condition has six droplets, and only 15% of all droplets have all their properties measured, while others are partially measured. The unmeasured properties are predicted by regression models (linear regression with various losses, boosting) for the next BO run. Finally, updated datasets corresponding to each target are combined for further machine learning tasks. The histogram and boxplot of targets during batch BO runs are given in (b).*

The combination of the batch BO step and the regression step forms our basic data generation workflow (Figure 2 (a)). We begin with a small initial dataset, called Run 0, chosen using a binary combination of P3HT(x%), ranging from (15% to 90%), with the rest of the mixture provided by a single CNT type solution (100-x)%. From this initial set, data generation proceeds with four workflows corresponding to different targets - minimizing and maximizing R (absorbance ratio) and Y (film conductivity), respectively, to explore configurations of sufficiently varying conductivities and reduce sampling bias towards high conductivity - implemented in parallel. The four collected datasets are combined at the end of the runs to form our full dataset for subsequent machine learning tasks. The histogram and boxplot of targets during batch BO runs are given in Figure 2(b), which shows the targeted exploration, as desired. As the BO appears to have converged after very few iterations, we have to make sure that the parameter space has been adequately explored. To this end, we use a teacher-student framework to confirm that the fast convergence is due to the good initial sample spread (See SI , section 3.4 for more details).

## 2.3 Graph-based model for CNT composition-property relationships

In this section, we outline our design of data-driven predictive models for the relationships between CNT composition and properties. The necessity to develop novel methods comes from two challenges that commonly arise in experimental settings. First, there are multiple measurements produced, many of which are intermediate quantities (e.g. thickness, absorbance ratio). They are not the final quantity of interest (i.e. conductivity), but they provide valuable information about the underlying system and our data driven model should be adapted to fully exploit them. Second, due to the large amount of noise present in experimental measurements, our regression model should also capture the uncertainty aspect in order to make meaningful and reliable predictions, and to perform inverse design. We will solve both of these problems via the development of a graph based regression model that actively accounts for uncertainty in predictions. Such models can be used in several ways, including property prediction, inverse design, and visualization of the "landscape" of conductivity with respect to the input composition, thereby deepening our understanding of the physical system at hand. Figure 3 gives an overview of the machine learning method, which we now discuss in detail.

### 2.3.1 Overall approach

Different from the traditional regression from the CNT composition C to conductivity $\sigma$ or Y, the graphical model takes the multiple outputs (A, R, and S) into account to predict Y, and pursues not only the mean accuracy but also the uncertainty of the predictions because of the inevitable error in experiments. Describing the CNT content and each measured quantity as nodes, and linking related nodes by directed edges, we can get a graph which characterizes the relations between CNT properties. Since our focus is on conductivity ($\sigma$ or Y), we consider directed acyclic graphs that have conductivity as its unique output (terminal node). Therefore, the graph $G$ can be regarded as a structured composite function approximating the conductivity. An example of this is $Y = f_{G,\mathcal{D},\mathcal{M}}(C) = F(C, R(C))$ corresponding to the graph $G_{C,1}^{(3)}$ in Figure3, where the functions $R(C)$ and $F(C,R)$ are parameterized functions by sub-models $\mathcal{M}$ to be learned from the dataset $\mathcal{D}$. We train a collection of such graph models (with different connectivity structures), from which we select a final model using a scoring system that takes into account both accuracy and uncertainty. Let us outline this approach in detail.

### 2.3.2 Modelling and learning under uncertainty

Although the film compositions are determined by user-defined experimental inputs, the nature of drop-casting introduces uncontrollable factors affecting the measured values, such as, inherent characterization noise and the variance in film quality. These factors suggest regarding the parameters as random variables. For $\sigma = 1/(R_s t)$, the noise is inherent in the sheet resistance and film thickness, and after taking a natural logarithm on the parameters, we can assume the noise of the logarithm follows Gaussian normal distributions (see SI section 3.1 for justifications for this assumption). Given the large range of conductivities that are accessible through the screening process, we model the logarithm of conductivity as,

$$Y_i \sim P_0(Y|C_i) = \mathcal{N}(\bar{Y}_i, \varepsilon_i^2),$$

where $Y_i = \log(\sigma_i)$, $\bar{Y}_i$ is the (unknown) conditional mean of $Y_i$ given $C_i$ and $\varepsilon_i$ is the noise level, which is estimated to be 0.3 from data. To accommodate the uncertainties in $Y$, we introduce two types of randomness into the graph-based model: random resampling of the training dataset from the original dataset $\mathcal{D}$ and random training procedure of sub-models $\mathcal{M}$. Therefore, the output of a graph $G$ describes a distribution $P$:

$$\hat{Y}_i \sim P(Y|C_i; G, \mathcal{D}, \mathcal{M}),$$

which is expected to approximate the distribution $P_0(Y|C_i)$. The difference between $P$ and $P_0$ is minimized by model selection over various graphical models. Randomness in resampling is realized by a subsampling strategy that is designed to take the imbalance in the distribution of $\mathcal{D}$ into account. Such imbalance arises because we are using BO to generate datasets, which causes more data to be sampled in regions with extreme objective values. General treatment for imbalanced datasets can be found in Torgo *et al*, [35], Branco *et al*, 2018[36], and the details of our resampling strategy is given in SI, section 3.2. Randomness in training procedure is realized by randomly choosing initial values and randomly searching hyper-parameters of machine learning models. Concretely, the sub-models $\mathcal{M}$ in graph $G$ are chosen from linear regression, Huber regression, and gradient-boosting models by usual cross-validation on the mean accuracy. Other models (including fully connected neural networks) are tested, but they are found via cross-validation to perform slightly worse, and hence are not included. We emphasize that both the randomness in the dataset and the training procedure are used to induce noise in our predictions, but they are not used to explicitly model the actual noise distribution of the experimental measurements. Rather, once a variety of noise is induced in our predictive models, we can then design a model selection strategy that selects models based both on its ability to perform accurate predictions and its ability to capture the uncertainties in experimental measurements. This is outlined below.

### 2.3.3 Graph model selection strategy

With the dataset $\mathcal{D}$ and the training procedure of sub-models $\mathcal{M}$ fixed, the final model $P(Y|C_i; G, \mathcal{D}, \mathcal{M})$ depends only on graph $G$, and the prediction performance can be measured according to the difference between this prediction distribution $P(Y|C_i)$ and the target distribution $P_0(Y|C_i)$. A good graphical model $G$ should have a high similarity or low distance between $P$ and $P_0$. Here, we estimate this distance by two scores, namely the R2-score and KL-score, using the mean and variance of $P$ and $P_0$ . The R2-score (only use the mean of $P$ and $P_0$) is widely used as a metric for regression accuracy, and the KL-score (use both the mean and variance of $P$ and $P_0$, and assume $P$ and $P_0$ are Gaussian normal distributions), named after the

KL-divergence, is a measure of distance (strictly, divergence) between two probability distributions. For the graphical model selection step, we use the R2-score as the main score (the higher the better, for mean accuracy) and KL-score (KL-divergence between $P(Y|C_i)$ and $P_0(Y|C_i)$, the lower the better, for uncertainty) as the secondary score of the graph models. The details of the definition and typical examples of the KL-score are given in SI, section 3.3.

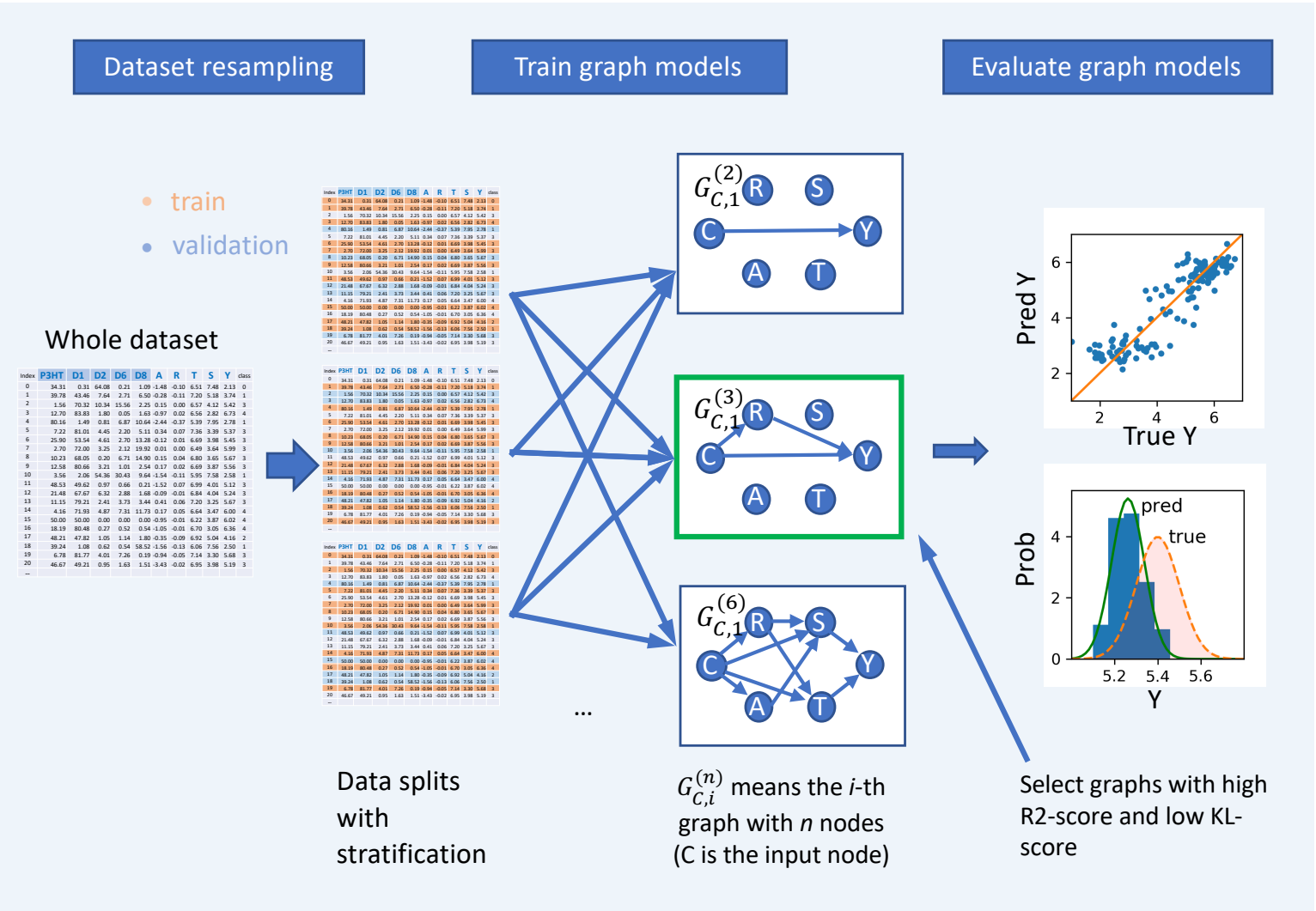

| Model selection | | | | |
|---|---|---|---|---|
| **Name** | **Inputs** | **R2 score (higher=better)** | **KL score (lower=better)** | **Graph** |
| $G_{C,1}^{(2)}$ | C | 0.808 | 64.9 | C→Y |
| $G_{C,1}^{(3)}$ | C | **0.824** | **33.0** | **C->R, CR->Y** |
| $G_{C,2}^{(3)}$ | C | 0.812 | 30.8 | C->S, CS->Y |
| $G_{C,1}^{(4)}$ | C | 0.824 | 27.1 | C->AR, CAR->Y |
| $G_{C,2}^{(4)}$ | C | 0.810 | 30.7 | C->RS, CRS->Y |
| $G_{C,1}^{(5)}$ | C | 0.825 | 27.8 | C->RS, S->A, CAR->Y |
| $G_{C,1}^{(6)}$ | C | 0.808 | 13.5 | C->AR, CAR->S, CR->T, ST->Y |
| $G_{CA,1}^{(3)}$ | CA | 0.814 | 34.6 | CA->Y |
| $G_{CA,1}^{(4)}$ | CA | **0.831** | 31.2 | CA->R, CR->Y |
| $G_{CR,1}^{(3)}$ | CR | 0.838 | 52.0 | CR->Y |
| $G_{CR,1}^{(4)}$ | CR | 0.841 | 49.7 | CR->S, CRS->Y |
| $G_{CR,1}^{(5)}$ | CR | **0.845** | 39.8 | R->A, CAR->S, CRS->Y |
| $G_{CS,1}^{(3)}$ | CS | 0.927 | 44.1 | CS->Y |
| $G_{CS,1}^{(5)}$ | CS | **0.931** | 18.22 | CS->T, ST->Y |
| $G_{CRS,1}^{(4)}$ | CRS | 0.927 | 32.7 | CRS->Y |
| $G_{CRS,1}^{(6)}$ | CRS | **0.930** | 17.6 | CRS->T, ST->Y |
| $G_{CARS,1}^{(5)}$ | CARS | 0.929 | 32.2 | CARS->Y |
| $G_{CARS,1}^{(6)}$ | CARS | **0.934** | 15.2 | CARS->T, ST->Y |

*Figure 3: Workflow of the graph-based machine learning method. (1) The graph-based regression model is used to predict the target 'Y' and its uncertainty. A graph model, e.g. $G_{C,1}^{(3)}$, is composited by some sub-regression-models, e.g. 'C->R' and 'CR->Y'. The sub-models are trained 20 times by traditional method (linear regression and GB with cross-validation) using randomly chosen training datasets. During the inference phase, the sub-models (randomly chosen one from the 20 sub-models) are composited according to the graph architecture, while the R2-score measures the prediction accuracy. The whole dataset is randomly resampled and split to training and validation datasets, where the randomness of the training dataset is expected to capture the uncertainty of 'Y', where the KL-score measures the discrepancy between prediction and true uncertainties. (2) The R2-score and KL-score are used to select graphs which have high accuracy and low discrepancy. (3) The selected graphs are used to extract the relationship between different properties and the conductivities. (4) The scores of some selected graphs are given.*

## 2.3.4 Results

The example graphs $G_{C,i}^{(n)}$ shown in Figure 3 are those having the CNT composition as the only input, which does not require additional measurements as inputs. This special setting could provide a tool to investigate the dependence of conductivity ($\sigma$ or Y) on the CNT composition (C). Our results in Figure 3 (inset Table) show that, to get predictions from C to Y, the simplest graph model 'C->Y' gives a good R2-score (0.808) but high KL-score (64.9), while the graph 'C->R, CR->Y', could give higher R2-score(0.824) and lower KL-score (33.0), which means the latter graph model captures the uncertainty of Y much better,. This performance improvement comes from the use of additional information of 'A, R, S, T' other than the conductivity itself. Noting that adding 'A, S, T' could further increase the R2-score and decrease the KL-score, but the improvements are small and the graphs are more complex. Therefore, we design a weighted score capturing accuracy, uncertainty, and graph complexity simultaneously (See SI, section 3.1 for more details). The graph model that has the highest weighted score, 'C->R, CR->Y' is chosen as our final predictive model using C as inputs.

We can add more input nodes in the graphical model if their values are measured experimentally. For example, we can use a measured R to help predict Y, instead of an R that is itself predicted from C using a regression model. This greatly expands the number of possible acyclic graphs to consider. In this case, our numerical results indicate that the prediction performance can be improved. The results in Figure 3 (inset Table) show that adding one node of 'A', 'R', and 'S' will improve the R2-score to 0.831, 0.845, and 0.931, respectively (the KL-score are also slightly improved). Therefore, we conclude that the most important node (disregarding thickness, since it is the most time-consuming step of the experimental workflow) to predict 'Y' is the sheet-resistance (S), followed by the absorption-ratio (R) and absorbance (A), i.e., 'S>>R>A', which is sorted by prediction improvements after the feature is taken into account. Using this order, we can further reduce experimental costs by mainly measuring the most important quantities. As a result, for the drop-casted workflow, the composition space can be effectively sampled by directly measuring the sheet resistance and final validation of the conductivity of the best performing samples can be performed by measuring the thickness of a chosen composition.

## 3 High-fidelity experiments based on data generation

We demonstrated that after 12 steps of batch BO, the maximum conductivity has no further improvement, which indicates that the fast convergence to locate the maximum conductivity as well as the optimal content region. To verify this fast convergence indirectly, we design an artificial ground truth CNT experiment, based on the graph-model, and then use the BO (same hyper-parameters with real experiments) to optimize film composition to maximize conductivities. Numerical results indicate that 7 BO steps are enough to locate the optimal region in parameter space for this artificial ground truth, with additional steps (10 more) having little improvement (for more details, refer SI, section 3.4). Since the artificial ground truth approximates the true target generating our dataset, it is likely that we have found the optimal regions in our experiments as well. Looking at the feature importances used in predicting electrical conductivity for both linear and gradient-boosting methods in the C → Y regression model, we noted that the top two features are l-SWCNT and DWCNT. Thus, we plot a full experimental manifold represented in a reduced two-dimensional plot comprising only these two CNT types (Figure 4b). Interestingly, we note a local maximum in the manifold emerging at ~10% of DWCNTs in a $L_{50}D_{10}P_{40}$ composite, which is unexpected from percolative electrical transport in these hybrids. In order to explore with finer composition resolution, we run additional validation experiments near this expected optimum and find that, within experimental error bar, the optimum is indeed observable, seen as red dots in Figure 4(c).

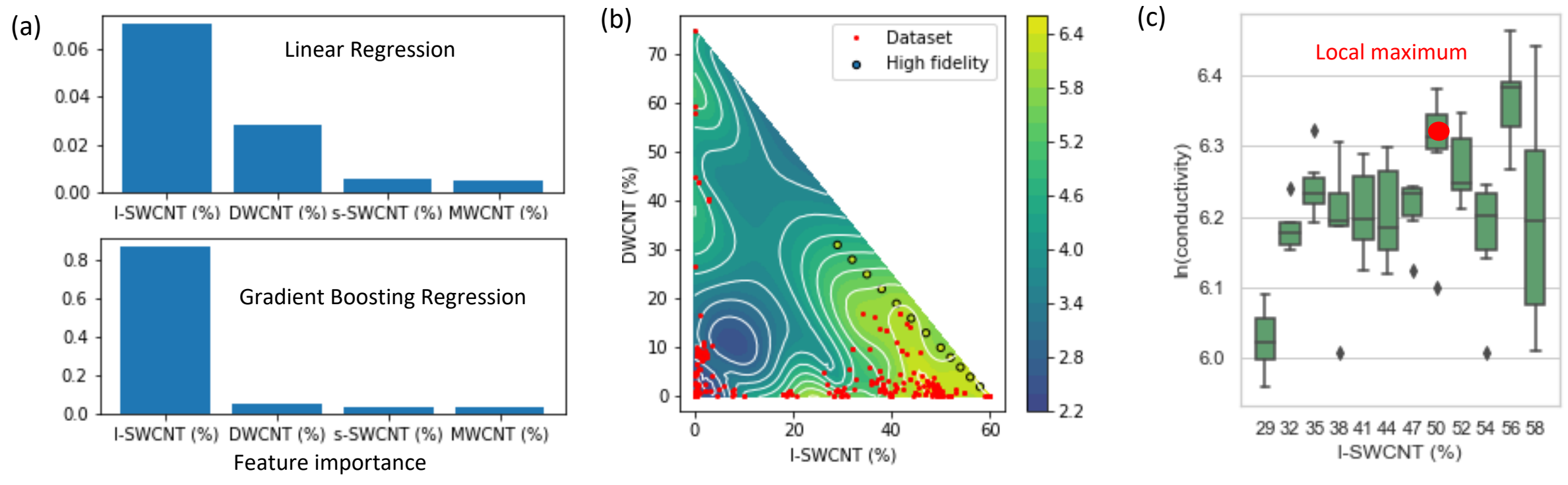

*Figure 4 (a) Feature importance of linear and gradient-boosting C → Y regression model. The top two features, l-SWCNT and DWCNT, are used to vituraize the experimental conductivity surface in (b). (b) Experimental manifold with the log of the electrical conductivity (Y) represented in the colored axis, as function of l-SWCNTs and DWCNTs compositions. Red dots are projected contents of the experiment dataset, and countour curves are provided by Gaussian-process-regression (main part of BO) with completed dataset as inputs. The circle dots are additional high fidelity experiments, to explore local maximum with respect to l-SWCNT and DWCNT mixtures, corresponding to measued Y in (c). (c) Boxplot of Y shows a local maximum at ~10% of DWCNTs in L50D10P40 composite.*

In order to interpret the ML suggestions and understand the mechanism behind composite electrical conductivity, we select four samples for comprehensive high-fidelity experimental analysis to relate the influence of P3HT/CNT composition ratio to electrical conductivity and absorption ratio. The samples include two high performing samples: $L_{60}P_{40}$, $L_{50}D_{10}P_{40}$, $L_{10}P_{90}$, a moderate performance composite (90 % P3HT and 10 % long single wall CNT, and $S_{60}P_{40}$, a low performance composite (with 40 % P3HT and 60 % short single wall CNT). The electrical conductivities of the four samples are 825 S/cm, 740 S/cm, 200 S/cm and 15 S/cm for $L_{60}P_{40}$, $L_{50}D_{10}P_{40}$, $L_{10}P_{90}$, and $S_{60}P_{40}$ respectively. The high-fidelity measurements include absorbance spectroscopy (from UV to mid IR range) to analyze the film crystallinity and polaron delocalization, Raman spectroscopy to understand the interaction between the P3HT and CNTs, and scanning electron microscopy (SEM) to evaluate the film morphology.

In the absorbance spectra, the π-π interaction of P3HT and the polaron delocalization length representing the film crystallinity can be observed by monitoring the red or blue spectral shift in the visible region for undoped films and Mid-IR region for $FeCl_3$ doped film respectively (shown in figure 5 (a) & (b)). Pristine and doped form of pure P3HT film is also measured for reference. All the spectra shown in figure 5 (a) & (b) are background subtracted, normalized and the Y-axis is offset in order to see the spectral features clearly. The absorption spectra of un-doped $L_{60}P_{40}$, and $L_{50}D_{10}P_{40}$ in the visible region (shown in figure 5 (a)) clearly shows red-shifted spectral features compared to pristine P3HT where the spectral feature due to π-π* transition and π-π interactions of well aggregated pure polymer chains appear at around ~ 2.36 and 2.06 eV respectively[37]. In addition to the red shifted spectral features, both $L_{60}P_{40}$, and $L_{50}D_{10}P_{40}$ films show fine features due to inter-band transition between van Hove singularities of single wall CNTs[38–40]. The red spectral shift due to P3HT aggregation and absorption features due to single wall CNT in both $L_{60}P_{40}$, and $L_{50}D_{10}P_{40}$ films indicate (1) efficient P3HT wrapping and ordering around the well-inter connected CNT network and (2) the availability of more electronic states due to higher CNT concentration contributes to the better electrical conductivity in these two composites compared to other films. In direct contrast, in the low electrical conductivity $S_{60}P_{40}$ sample, the spectral features due to π-π interaction are weak and shifted towards blue compared to P3HT, which indicates that the introduction of short single wall carbon nanotubes is not contributing to the alignment of polymer chains. The CNT fine features are also not seen in $S_{60}P_{40}$ although the percentage of CNT is 60% indicating unfavorable polymer wrapping around CNT. Both pristine P3HT and $L_{10}P_{90}$ show spectral features corresponding to well-aligned P3HT where the features due to intrachain (π-π*) and interchain interaction (π-π) are seen clearly (Refer Figure S8 for the complete UV-VIS-MIR spectra of all the pristine and doped films).

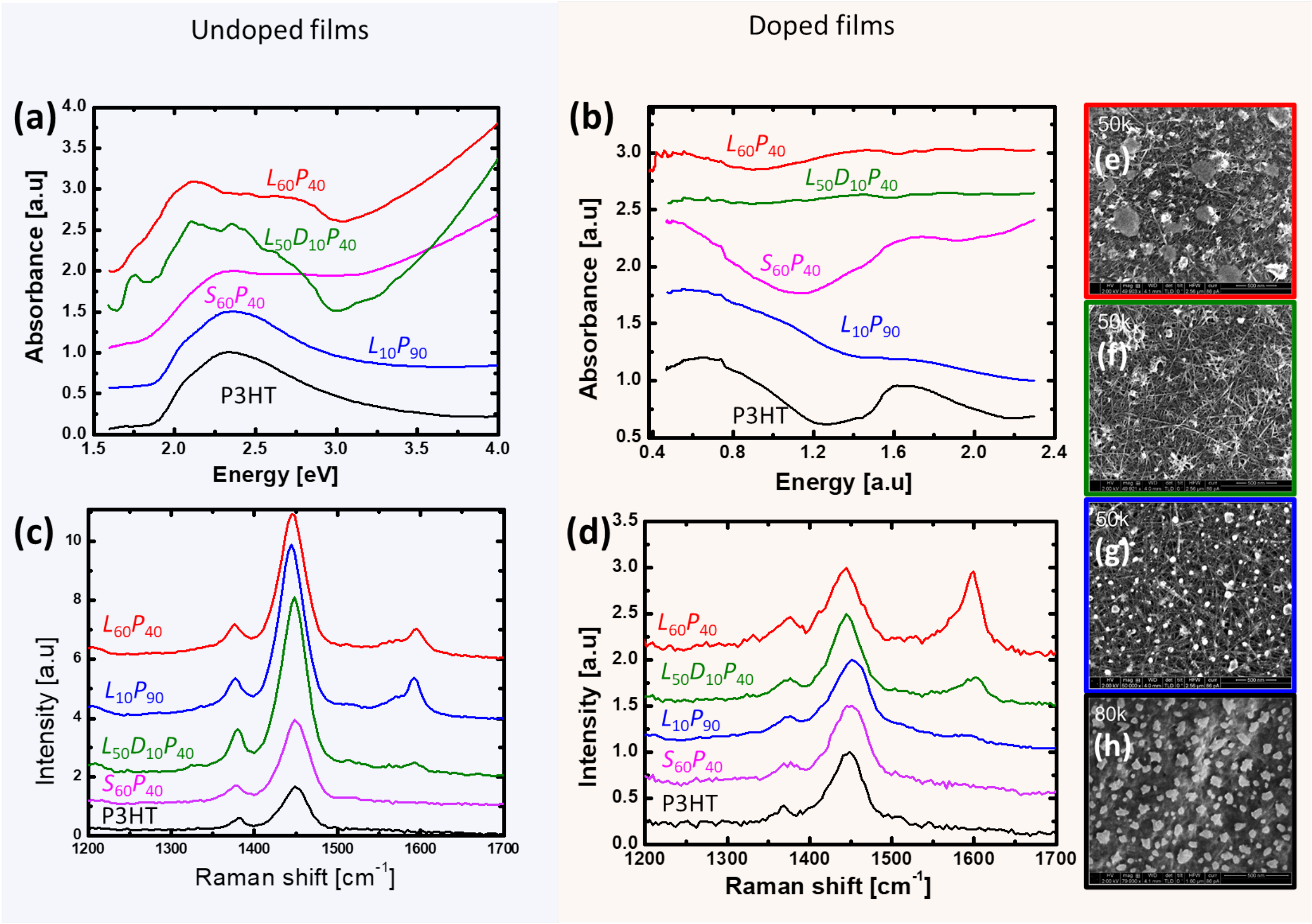

*Figure 5: Optical and Morphological investigations on selected samples for interpretable ML model. (a): Absorbance spectra of un-doped P3HT/CNT composite films showing more red shift of absorption shoulder due to π-π interaction accompanied with fine absorption features due to CNT in both* $L_{60}P_{40}$ and $L_{50}D_{10}P_{40}$ *films compared to other composites; (b) Absorbance spectra of doped films showing the position of P1 band due to polaron delocalization; (c) Raman spectra of un-doped and (d) doped films* showing *the* C=C skeletal stretching vibration accompanied with 'G' band vibration due to CNT and (e-h) showing the SEM images of doped $L_{60}P_{40}$, $L_{50}D_{10}P_{40}$, $L_{10}P_{90}$, and P3HT films respectively.

The absorbance spectra of $FeCl_3$ doped films are shown in figure 5(b). Doping of P3HT with $FeCl_3$ induces polaron formation that introduces two localized electronic states in the bandgap. The optical transition between these two levels create two additional absorption bands at wavelengths longer than the neutral excitons found in undoped films[41]. The high energy polaron band in $FeCl_3$ doped P3HT appears at around 1.6 eV ($P_2$ band) and the low energy polaron band appears at around 0.5 eV ($P_1$ band). Generally speaking, the $P_1$ band is isolated from other optical transitions and thus it can be used to monitor the extent of polaron delocalization. Highly ordered P3HT films doped with $FeCl_3$ have shown a distinct $P_1$ band at 0.38 eV[42], although the energy location and absorption strength depends on the degree of crystallinity. The absorption spectra shown in Figure 5(b) is normalized to the low energy band at around 0.5 eV. The $P_1$ energy measured for P3HT and $L_{10}P_{90}$ is ~ 0.6 eV, that shifts towards 0.5 eV in $L_{60}P_{40}$, $L_{50}D_{10}P_{40}$, and $S_{60}P_{40}$. It can be seen that the intensity of $P_1$ band in $L_{10}P_{90}$ is much stronger and broader than the other composites indicating the wide distribution of polaron delocalization with much stronger degree of crystallinity, but its electrical conductivity is lower than $L_{60}P_{40}$ or $L_{50}D_{10}P_{40}$, indicating that the interconnectedness of the carbon nanotubes is the dominant factor affecting the electrical conductivity of the composite compared to the crystallinity of the P3HT electrical conductivity[40].

The Raman spectra of undoped and doped films using 532nm laser excitation are shown in figure 5 (c) and (d). In this analysis, we focus on the prominent vibrational modes in the wavenumber region between 1300 and 1700 $cm^{-1}$ as it explains the interaction between P3HT and CNT. The pristine P3HT shows two prominent vibrational peaks at 1382 $cm^{-1}$ and 1450 $cm^{-1}$ that are assigned to C-C intra-ring and symmetric C=C stretching vibrations, consistent with literature values[43]. The Raman spectra of $L_{10}P_{90}$ shows three peaks at 1380 $cm^{-1}$, 1447 $cm^{-1}$ and 1593 $cm^{-1}$, which are attributed to C-C intra-ring stretching, C=C skeletal stretching of thiophene rings and characteristic 'G' band of SWNTs due to in-plane stretching of $E_{2g}$ mode[44]. The Raman spectra of $L_{60}P_{40}$, and $L_{50}D_{10}P_{40}$ shows the characteristic peaks of P3HT and CNT at around 1378, 1445 and 1592, however $S_{60}P_{40}$, which has 60 % short single short single wall carbon nanotube does not show the characteristic 'G' band signal indicating inhomogeneous distribution of carbon nanotubes. The Raman spectra of pristine CNTs are shown in Figure S9.

The Raman spectra of $FeCl_3$ doped films are shown in figure 5 (d). The presence of delocalized polarons shifts the symmetric C=C stretching vibrations to lower wavenumber compared to the undoped films as the presence of polarons weakens the bond strength, shifting the stretching to lower energy modes. The C=C stretching vibrations in both doped $L_{60}P_{40}$, and $L_{50}D_{10}P_{40}$ films exhibit larger shift towards the lower wavenumber compared to the other composites in addition to the strong 'G' band contribution indicating the better degree of polaron delocalization and efficient polymer wrapping in a well-connected CNT network, which is consistent with the absorbance result shown in figure 5(b). Therefore, absorbance and Raman spectra allow us to conclude that the better degree of polaron delocalization and well-connected CNT network with a high density of mobile charges contribute to the observed better electrical conductivity in $L_{60}P_{40}$, and $L_{50}D_{10}P_{40}$ films.

To further evaluate the influence of morphology on electrical conductivity, the Scanning Electron Microsopy (SEM) of doped films, $L_{60}P_{40}$, $L_{50}D_{10}P_{40}$, $L_{10}P_{90,}$ and P3HT, are obtained as shown in figure 5 (e-h) respectively. (Refer Figure S10 for the SEM imagecof different magnifications) The SEM image of all the three composite films (e-f) shows a homogeneous distribution of CNT wrapped with polymer; the well-interconnected CNT networks with 10 % long single wall CNT concentration is shown in figure 5 (f). The higher concentration of long single wall CNTs in both $L_{50}D_{10}P_{40}$ and $L_{60}P_{40}$ films is expected to be the differentiating factor for the higher electrical conductivity compared to the $L_{10}P_{90}$ film, where a higher number of mobile charges due to doped CNTs is present in the film, thereby modulating the band structure. The SEM image of $S_{60}P_{40}$ can't be measured which is consistent with the spectroscopy and low electrical conductivity where polymer wrapping in the short single wall CNT is not favorable for electrical conduction.

# 4 Discussion – beyond validation and broader applications of this methodology

We would like to emphasize the inroads we have made into application of machine learning techniques to small, but well-labelled data, common in materials science. In classical regression models, the regression function is trained from the dataset, only including the inputs and target values. In contrast, our graph-based model takes potentially measured proxy quantities, such as the absorbance ratio, into account; such practices can be generalized to other physical systems. The learned graph models could have better prediction accuracy and uncertainty estimation. This is useful to visualize the landscape of the target, decrease measurement cost, perform inverse design, and determine relevant features for prediction (feature selection). Let us illustrate these general advantages in the specific context of our CNT experiments.

First, we discuss how having a surrogate model can help visualize complex relationships between physical quantities. For example, during the prediction of 'Y' by using the graph 'C->R, CR->Y', we also predicted the value of 'R'. Figure 4(a) visualizes the landscape of 'Y' and the relationship between 'R' and 'Y' on a subspace of compositions with only l-SWCNT and DWCNT. Using the predicted values locally, we can fit a polynomial function to smooth the landscape. For the R-Y relationship, a bilinear function could fit well, which suggests a local bilinear relationship between R and Y. This indicates that once the P3HT concentration is fixed, the ratios of carbon nanotubes determine the conductivity. This is a clear trend that is consistent with percolation theory, as the CNTs allow for delocalization of charges, and are effectively responsible for how well charges flow in the hybrid system. Therefore, for each ratio of carbon nanotubes, the P3HT plays a secondary role to hold the hybrid together, but ultimately the degree of charge transfer is controlled by the relative mixing of different CNTs.

Next, the fact that a graph model can handle multiple inputs and outputs can help with data imputation. During data generation, we used the simple linear regression and gradient boosting model (C→Y) to predict the conductivity. However, our graph model results show that we can leave some quantities (especially the thickness 'T', whose measurement is the most time-consuming) to be unmeasured, and the learned graph model could predict 'Y' well. Therefore, the measurement costs in future experiments can be reduced. Our graph-based model also provides a tool to determine important quantities to predict a given target and detect the relationships between them.

Finally, the graph based machine learning workflow introduced in this paper forms a general approach to build models amongst multiple noisy inputs and outputs, especially when there is no *a priori* known relationships between them. This is a common occurance when performing machine learning on experimental data. Our method is not only a regression model (with a fixed graph) but also a model/feature selection strategy considering both accuracy and uncertainty, with the eventual task of determining an optimal graphical model on which to build our regressor. Such an approach has resemblance to classical methods in multi-target regressions (MTR)[45–48], probabilistic graphical models (PGM)[49], and uncertainty estimation learning[50,51] However, our problem setting is sufficiently different from usual statistical applications, and necessitates the development of novel techniques (Refer SI, section 3.5 for detailed discussion).

# 5 Conclusion

We have demonstrated a platform for machine learning-assisted high-throughput exploration and optimization for functional composites consisting of Regio-Regular poly-3-hexylthiophene (P3HT) and carbon nanotubes (CNTs). The platform includes an automated flow system, drop casting facility, Hyperspectral imaging system and four-point probe for fast material processing and optical/electrical diagnostics. With the combination of this semi-automated high-throughput platform and machine learning in the loop, we showed rapid optimization of 5-dimensional composition space, achieving state-of-the-art electrical conductivity as high as 1200 S/cm. Furthermore, the graph based machine learning methodology developed here may be applied to a wide variety of problems involving noisy measurements of several quantities, where model selection and regression must be coupled in a principled way to achieve accurate and robust predictions.

**Supplementary Information**

**Machine learning and high-throughput methodology for robust design of P3HT-CNT composites for high electrical conductivity**

Daniil Bash*[1,2], Cai Yongqiang*[3,4], Vijila Chellappan*[1], Wong Swee Liang[1], Xu Yang[1], Pawan Kumar[1], Tan Jin Da[1,5], Anas Abutaha[1,6], Jayce Cheng[1], Lim Yee Fun[1], Siyu Tian[7], Danny Zekun Ren[7], Flore Mekki-Barrada[8], Wai Kuan Wong[8], Jatin Kumar[1], Saif Khan[8], Li Qianxiao[#3,9], Tonio Buonassisi[#7,10], Kedar Hippalgaonkar[#1,11]

*equal contribution, correspondence to qianxiao@nus.edu.sg , buonassisi@mit.edu, kedar@ntu.edu.sg

[1] Institute of Materials Research & Engineering, Agency for Science Technology and Research, 2 Fusionopolis Way, 138634 Singapore
[2] Department of Chemistry, 3 Science Drive 3, National University of Singapore, 117543 Singapore
[3] Department of Mathematics, National University of Singapore, 21 Lower Kent Ridge Rd, 119077 Singapore
[4] School of Mathematical Sciences, Laboratory of Mathematics and Complex Systems, MOE, Beijing Normal University, 100875 Beijing, China
[5] NUS Graduate School for Integrative Sciences & Engineering (NGS), University Hall, Tan Chin Tuan Wing Level 04, #04-02, 21 Lower Kent Ridge Road, 119077, Singapore
[6] Qatar Environment and Energy Research Institute, Hamad Bin Khalifa University, Qatar Foundation, Doha, 34110, Qatar
[7] Singapore-MIT Alliance for Research and Technology SMART, 138602, Singapore
[8] Department of Chemical and Biomolecular Engineering, National University of Singapore, 117585, Singapore.
[9] Institute of High Performance Computing, 1 Fusionopolis Way, #16-16, Connexis 138632, Singapore
[10] Massachusetts Institute of Technology, Cambridge, MA 02139, USA.
[11] Department of Materials Science and Engineering, Nanyang Technological University, Singapore 639798, Singapore

# 1. Materials & Sample Preparation

## 1.1. Materials

All materials were used as purchased without purification, unless otherwise stated. Anhydrous 99% pure o-DCB was purchased from Sigma Aldrich. 98% pure long- and short-SWCNTs were purchased from US-Nano, while 99% pure MWCNTs and 98% pure DWCNTs (60% DWCNTs specifically, but 98% nanotubes) were purchased from Nano Integris. Electronics grade P3HT flakes, obtained from Reike Metals and anhydrous electronic-grade $FeCl_3$, purchased from Sigma Aldrich, were stored under inert atmosphere inside glovebox. Reagent grade nitromethane was purchased from Sigma Aldrich and was additionally dried and purified using procedure described elsewhere[1].

## 1.2. Preparation of CNT stock solutions

The procedure for dispersion of CNTs was based on work by Bar-Hen et al[2]

### 5.1 L60P40

To 18 mg threads of long SWCNTs in 50 ml V-shaped flask 1.2 ml of 10 mg/ml solution of P3HT in o-DCB was added and filled to 30 ml with o-DCB. This mixture was then sonicated at 100W in pulse mode (59 seconds 'ON' & 15 seconds 'OFF') with a sonicator probe for 3 hours. The resulting dispersion was transferred to a 50 ml falcon tube and centrifuged at 7500 rpm for 5 minutes. Then the dispersion was gently transferred to a storage vial for later use, while the sediment was transferred to a small vial, dried out over few days and weighted afterwards.

### 5.2 S85P15

To 25.5 mg powder of short SWCNTs in 50 ml V-shaped flask 450 ul of 10 mg/ml solution of P3HT in o-DCB was added and filled to 30 ml with o-DCB. This mixture was then sonicated at 100W in pulse mode 59 sec on 15 sec off with a sonicator probe for 2 hours. The resulting dispersion was transferred to a 50 ml falcon tube and centrifuged at 5000 rpm for 5 minutes. Then the dispersion was gently transferred to a storage vial for later use, while the sediment was transferred to a small vial, dried out over few days and weighted afterwards.

### 5.3 M85P15

To 25.5 mg powder of MWCNTs in 50 ml V-shaped flask 450 ul of 10 mg/ml solution of P3HT in o-DCB was added and filled to 30 ml with o-DCB. This mixture was then sonicated at 100W in pulse mode 59 sec on 15 sec off with a sonicator probe for 2 hours. The resulting dispersion was transferred to a 50 ml falcon tube and centrifuged at 5000 rpm for 5 minutes. Then the dispersion was gently transferred to a storage vial for later use, while the sediment was transferred to a small vial, dried out over few days and weighted afterwards.

## 5.4 D75P25

To 22.5 mg powder of DWCNTs in 50 ml V-shaped flask 750 ul of 10 mg/ml solution of P3HT in o-DCB was added and filled to 30 ml with o-DCB. This mixture was then sonicated at 100W in pulse mode 59 sec on 15 sec off with a sonicator probe for 2 hours. The resulting dispersion was transferred to a 50 ml falcon tube and centrifuged at 5000 rpm for 5 minutes. Then the dispersion was gently transferred to a storage vial for later use, while the sediment was transferred to a small vial, dried out over few days and weighted afterwards.

**1.3 Preparation of P3HT/CNT mixtures and films**

All P3HT/CNT hybrid mixtures were prepared using microfluidic flow reactor in plug-flow mode. The reactor consists of 3 LabView-controlled dual syringe pumps, mixing junction, gas-flow controller to separate flow into plugs (slugs), and a refurbished 3D-printer stage. The internal diameter (ID) of tubing for the flow reactor was chosen to be 0.25 mm, as it was found to be optimal for the drop-casting process, ensured better mixing, and facilitated with plug formation.

The experimental conditions were used in form of Design of Experiments (DoE) file in .csv format, generated by the Bayesian Optimization algorithm. This file contained data for volumes of solutions to be mixed in the flow reactor, while combined flowrate was kept constant throughout the experiment. To ensure repeatability of the experiment, each experimental condition consisted of 5-6 identical samples, while first 10-20 droplets of every condition were discarded, to eliminate possibility of cross-contamination and accommodate for fluctuations of syringe pump flow due to change of individual flow rates.

Samples were drop-casted onto treated double-polished fused silica substrates. The substrates were subsequently sonicated for 5 mins in acetone bath, 5 mins in IPA bath, and treated for 10 mins in UV-ozone cleaner at 100°C. Substrates were used for drop-casting within 10 minutes from treatment. After drop-casting, samples were left to dry unmoved. Then dry samples were transferred for doping, detailed procedure for which can be found below.

**1.4.Doping of films**

The drop-casted films on 4” quartz wafer was doped using 0.03 M FeCl3 solution in anhydrous nitromethane, prepared with anhydrous electronic-grade FeCl3. The wafers were immersed in FeCl3 solution for 2-3 seconds. After immersion, excess solution was removed by blow-drying with nitrogen gun. Both preparation of doping solution and doping itself were carried in an inert atmosphere environment.

## 2. Measurements

### 2.1. Absorbance spectra using hyperspectral imaging (HSI) system

In this work, we use PIKA-L (Resonon, USA) HSI in transmittance mode to measure the optical absorbance of P3HT/CNT composite films. The wavelength range of the HSI system is from 400 nm to 1000 nm with spectral resolution of 2.1 nm and spatial resolution of ~120 μm. The measurement and analysis of the HSI image cube involves four main steps in the following order: 1. System calibration, 2. Recording of sample image 3. Processing the image cube and 4. Analyzing the spectral features. The system calibration involves the correction of the sample image from the dark current of the camera and the intensity profile of the light source. The dark response of the camera ($D$) is recorded by turning off the light source and covering the camera lens while the intensity profile of the light source is recorded by measuring the intensity of light transmitted through a reference quartz substrate ($I_0$). Upon completion of the system calibration, we scan the sample image that measures the intensity of the light transmitted through the sample ($I$). The transmittance ($t$) of the sample is calculated using the equation given below:

$$t = \frac{I - D}{I_0 - D} \qquad (1)$$

From the image cube recorded, the mean transmittance spectra of the region of interest (ROI) of every drop-casted film is obtained by averaging ~20 x 20 pixels in the middle of the droplets. The absorbance values ($A$) are then calculated using the equation as shown below:

$$A = -\log(t) \qquad (2)$$

The absorbance spectra are obtained by plotting the absorbance values against the wavelength in nanometers (nm).

**Absorbance spectral analysis of drop-casted P3HT/CNT composite films**

The example HSI cube recorded on P3HT/CNT composite films prepared by drop casting is shown in figure S1. There are 36 droplets with 6 unique conditions are drop casted on a quartz wafer.

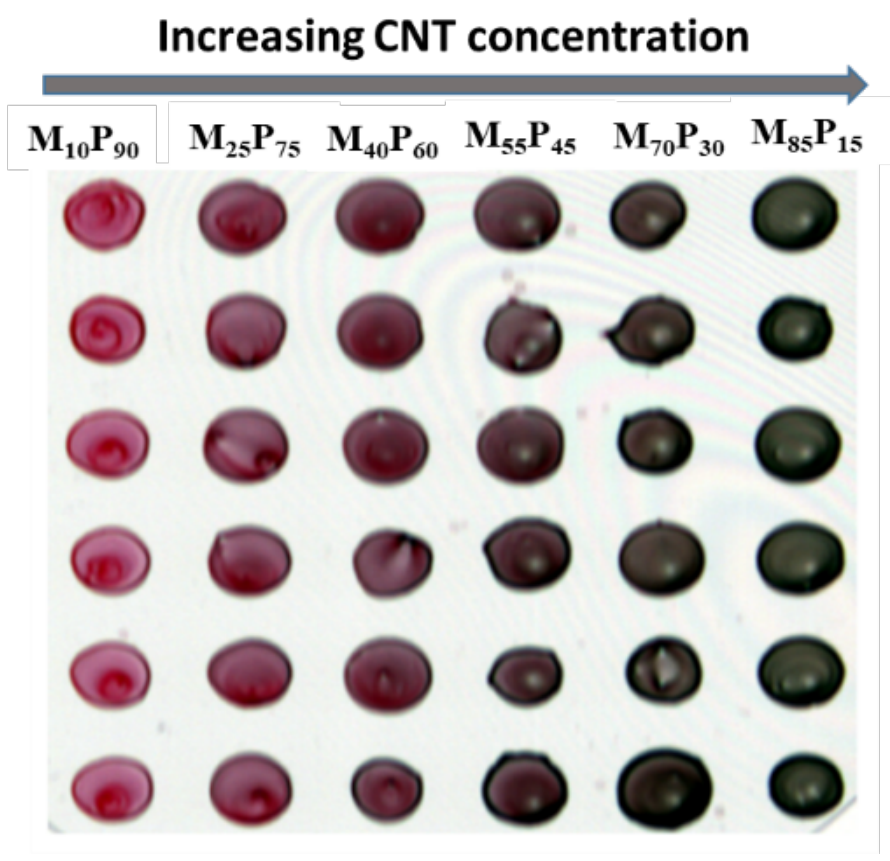

*Figure S1 Example HSI cube of P3HT/CNT composite film with varying multiwall CNT concentration*

Each column in Figure S1 represents a P3HT/CNT composition and the % CNT content increases from 10% to 85% in steps of 15% from the leftmost column to the rightmost column. There are 6 replicates of every P3HT/CNT composites in order to account the variation of film thickness due to drop casting. The averaged absorbance spectra from ~20x20 pixels in the middle of the droplets is shown in Figure S2

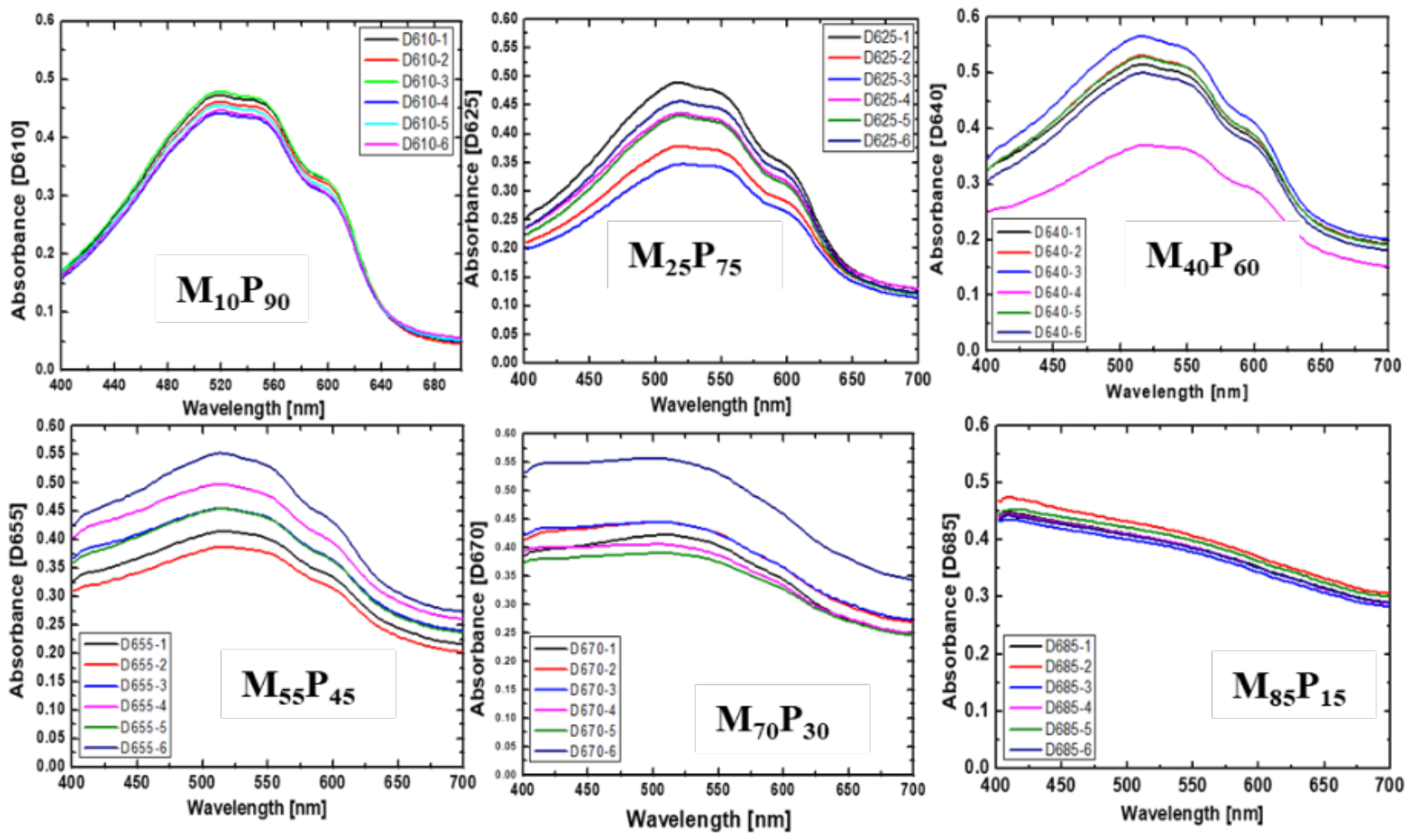

*Figure S1 Absorption spectra of all 36 films (one wafer) grouped according to their % CNT content.*

### 2.2. Four-point probe measurement

Sheet resistance of Fecl3 doped films are measured using semi-automated four-point probe (4pp) measurement setup. Keithley 2450 was used as a source meter to measure I-V curve for each droplet in a four-point probe configuration. From the I-V curve, the slope was calculated and multiplied by the geometry factor (3.8) to determine the sheet resistance of each droplet. The conductivity (σ) of each droplet was calculated using the following equation:

$$\sigma = \frac{1}{R_S t} \quad (3)$$

Where $R_S$ is the sheet resistance and *t* is the thickness of the drop casted film. The thickness of 15% droplets were measured using surface profilometer and the remaining values were inferred using machine learning (see below for detailed procedure).

### 2.3. Surface Profilometer

Surface profilometry was done on the KLA Tencor P-16+ profiler. Average thickness value, measured between imprints of inner two electrodes of 4pp, was used in the experiment. This was done for consistency of the measurements between experiments, and to ensure that thickness data is indexed to exactly same position as 4pp data. Error of the thickness measurement was recorded as fractional deviation of thicknesses on the edges of measured section from the average thickness.

## 3. Machine learning enabled data generation, model selection and property optimization

### 5.5 3.1 Data pre-processing

Before using the raw data collected from CNT experiments into machine learning, we need a pre-processing step, i.e. to examine the intrinsic properties. For our CNT dataset, the attributes are [absorbance], [absorption ratio], [sheet resistance], [film thickness] and [conductivity]. In Figure S2, we examine the noise levels of these attributes with and without taking the logarithm. It is obvious that some attributes such as the [sheet resistance] has a large noise as the value itself is large. After taking a logarithm on [sheet resistance], hereafter denoted by 'S', the noise becomes less correlated with the scale of 'S'. This is the case for the [absorption ratio], [sheet resistance], [film thickness] and [conductivity], but not the [absorbance]. Therefore, we will use the quantities listed in Table S1 for our forward data analysis processes.

Note that Figure S2 also indicates the noisy nature of the CNT experiments. To estimate the noise level for 'A', 'R', 'S' and 'T', we calculate the standard derivation of these attributions over different droplets and then average them over different CNT contents. The estimated noise levels are also listed in Table S1. Note that for 'T' and 'Y', only a few CNT samples have more than two measured droplets, therefore the noise level is roughly estimated. The noise level for 'Y' is estimated approximately 0.3 (since 'Y' is calculated from 'S' and 'T' whose noise level are estimated as 0.2 and 0.1, respectively). The noise level estimates are obtained for subsequent model selection strategies accounting for uncertainty.

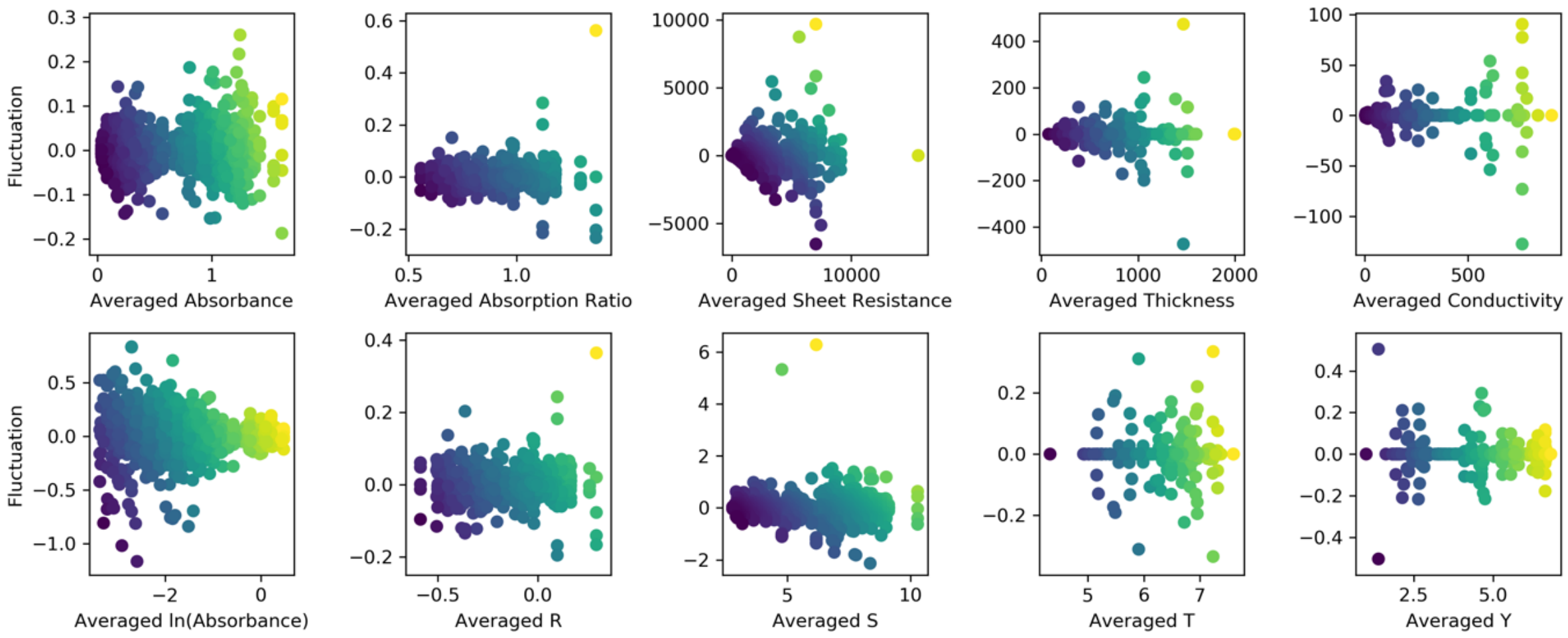

*Figure S2 The fluctuation for CNT attributes. The top row is for the original quantities, and the bottom is after the logarithm. The averaged values come from the six droplets that has the same CNT contents.*

*Table S1 Node and attributes*

| Node | Attributes | Noise level (esitimated) |
|---|---|---|
| C | CNT contents | - |
| A | Absorbance | 0.05 |
| R | Ln(Absorption ratio) | 0.05 |
| S | Ln(Sheet resistance) | 0.2 |
| T | Ln(Film thickness) | 0.1 |
| Y | Ln(Conductivity) | 0.3 |

## 5.6 3.2 Dataset resampling

Since our dataset $\mathcal{D}$ is collected via BO iterations that maximizes or minimizes (for the purpose of exploration) the absorption ratio and the conductivity, CNT contents are no longer uniformly sampled from the whole content simplex domain. The high/low conductivity regions have more data points of CNT contents than other regions. For example, in the Run1-12 data, the Euclidean distances from one content to its closest content have many small values. Figure S3 gives the boxplot of the distances (the median value is 2.51, mean is 3.32). Therefore, we treat our dataset as an imbalanced dataset when we want to use machine learning methods to discover the relations on the whole content space.

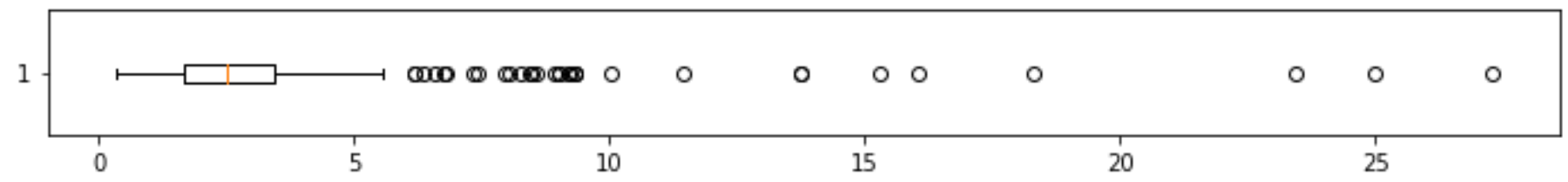

*Figure S3 Distance to the closest content*

There are many treatment for regression problems with imbalanced datasets[3,4]. Here we use the following resampling strategy:

1. Set a threshold distance $d_0 = 1.0$, and define a function $h(x) = e^{-x^2/2d_0^2}$;
2. For the $i$-th data point of CNT contents, calculate the distance to all data points (include the $i$-th data point itself), $\{d_{i1}, d_{i2}, \ldots\}$, and then assign a number $p_i = 1/(\sum_j h(d_{ij}))$ (a number between 0 and 1);
3. Keep or drop data points. Keep the $i$-th data points with probability $p_i$.
4. Choose the CNT droplet within a same data point. If the $i$-th data point is kept, then we randomly choose a CNT droplet in this content from all six droplets.

Note that $p_i$ will be small if the $i$-th data point has many close data points, and $p_i \approx 1/m_i$, where $m_i$ is the number of very close data points. Using this resampling strategy, we can get many randomly resampled datasets $\mathcal{D}_1, \mathcal{D}_2, \ldots$ which are used for subsequent machine learning procedures. Particularly, during the cross-validation steps, we group the data by the value of $Y$=ln(conductivity)._Since the range of $Y$ is $Y \in (0,7]$ (The histogram is given in Figure S4), we assign a group label for each value to stratify the train and test split of the dataset (Table S2).

*Table S2 Label of the data points*

| *Label* | *0* | *1* | *2* |
|---|---|---|---|
| $Y$ | $Y \leq 2.5$ | $Y \in (2.5,4]$ | $Y \in (4,5]$ |
| *Label* | *3* | *4* | |
| $Y$ | $Y \in (5,6]$ | $Y > 6$ | |

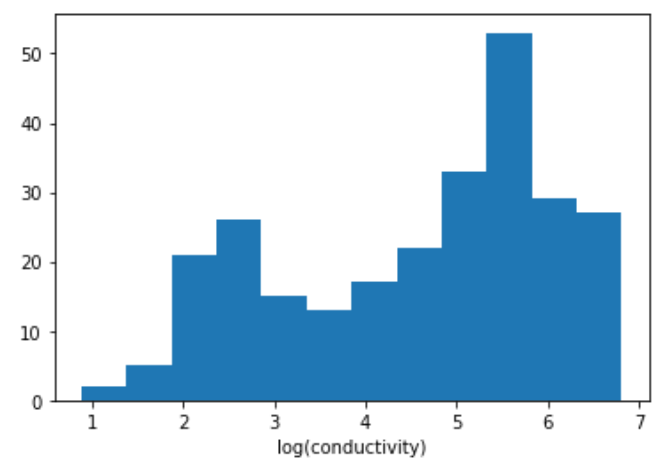

*Figure S4 Histogram of Y.*

## 5.7 3.3 Graph-based model

Here we use a directed graph to model the approximation of $f(C)$. For example, the 'C→R, CR→Y' graph means the function $f_1$:

$$f_1(C) = Y(C, R(C)),$$

where $R(C)$ and $Y(C,R)$ are functions learned from the dataset. The main feature of the graph-based approximation is the compositional structure, which can be regarded as a prior of the function $f(C)$.

**Candidate graphs** Given the nodes, and fixed inputs (C) and output (Y), there are a large number of candidate graphs which could be used to construct approximations of $f(C)$. Thus, we need some conditions to reduce the number of candidate graphs for search. We employ the following conditions: (1) the longest path in the graph contains at most four nodes, (2) if both S and T are in the graph then they are the only source to Y (since we know the exact relationship from S and T to Y). Then there are 368 candidate graphs and Figure S5 gives some examples.

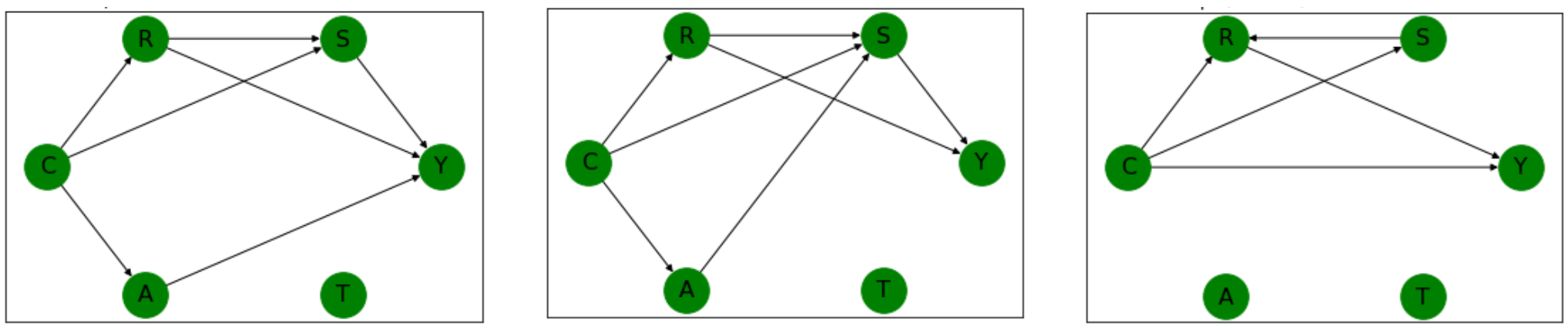

*Figure S5 Example of candidate graphs*

### Training graph models

Here we use the graph-based model to perform regression analysis on the CNT dataset. In each graph, the regressors of each node (except the input node C) are chosen from linear regression, Huber regression and gradient-boosting model determined by cross-validation. Since the dataset are obtained from the BO suggestions, we treat it as an imbalanced dataset and design a resampling strategy to generate training and test datasets. Due to the randomness of resampled training dataset, the prediction of a given graph model could be regarded as a random variable, and then the prediction performance can be quantified by the discrepancy between the prediction distribution and the true value distribution (because the measurement is noisy). Particularly, we use the r2-score (the higher the better) as the main score which measures the collinearity of the predicted and target mean values, and KL-score (the lower the better) as the secondary score of the graph models to measure how well the uncertainty in the data has been captured.

### Evaluating and selecting graph models

In the graph model selection part, we use the mean of $P(Y|C_i)$ as the prediction and use the r2-score as the **main** model score. (Note that other score functions have been tested, but r2-score was the most reliable one.) In addition, we will use the KL-divergence between $P(Y|C_i)$ and $P_0(Y|C_i)$ as the **secondary** score (the lower the better). The KL-divergence between two Gaussian distributions, $P(x) = \mathcal{N}(\mu, \sigma^2)$ and $P_0(x) = \mathcal{N}(\mu_0, \sigma_0^2)$, is given by:

$$\mathrm{KL}(P, P_0) = \int P(x) \ln \frac{P(x)}{P_0(x)} dx = \frac{1}{2}\left(\frac{\sigma^2 + (\mu - \mu_0)^2}{\sigma_0^2} - 1 - \ln \frac{\sigma^2}{\sigma_0^2}\right).$$

The r2-score is defined as:

$$\mathrm{r2score} = 1 - \frac{\sum_i \left(Y_i - \hat{Y}_i\right)^2}{\sum_i (Y_i - \bar{Y})^2},$$

For the KL-score, we use the following definition according to the KL-divergence of two Gaussian distributions:

$$KLscore_1(Y, \hat{Y}, \sigma, \hat{\sigma}) = \frac{1}{2}\left(\frac{\sigma^2 + \left(Y - \hat{Y}\right)^2}{\hat{\sigma}^2} - 1 - \ln \frac{\sigma^2}{\hat{\sigma}^2}\right),$$

$$KLscore_2\left(Y, \hat{Y}, \sigma, \hat{\sigma}\right) = \frac{1}{2}\left(\frac{\hat{\sigma}^2 + \left(Y - \hat{Y}\right)^2}{\sigma^2} - 1 - \ln \frac{\hat{\sigma}^2}{\sigma^2}\right),$$

where $\hat{Y}, \hat{\sigma}$ are the prediction of $Y$ and its standard derivation $\sigma$. Note that the KL-divergence between two distributions is asymmetric. The $KLscore_2$ is almost linearly related to r2-$score$, therefore we use $KLscore_1$ as the KL-score.

$$KLscore = \sum_i \min_{\sigma_0 \leq 0.3} KLscore_2(Y_i, \hat{Y}_i, \sigma_i, \hat{\sigma}_i).$$

To capture model accuracy, uncertainty, and graph complexity simultaneously, we combine the R2score and KL scores by a weighted score defined by:

$$score = 20 * r2score - 0.01 * \ KLscore - 0.1 * E,$$

where $E$ is the number of edges in a graph which characterize the complexity of the graph.

*Table S3 Typical example of KL scores*

| | $(Y,\hat{Y},\sigma,\hat{\sigma})$ | $(Y,\hat{Y},\sigma,\hat{\sigma})$ | $(Y,\hat{Y},\sigma,\hat{\sigma})$ | $(Y,\hat{Y},\sigma,\hat{\sigma})$ |
|---|---|---|---|---|
| | (6.0,6.5,0.3,0.01) | (6.0,6.5,0.1,0.0.01) | (6.0,5.0,0.3,0.3) | (6.0,4.0,0.1,0.1) |
| $KLscore_1$ | 1696.10 | 1297.20 | 5.56 | 200 |
| $KLscore_2$ | 4.29 | 14.31 | 5.56 | 200 |

## 5.8 3.4 Verification of the BO convergence

In our real experiment and BO search, the BO only ran for 12 steps. A natural question is do we need more steps? To obtain an indirect verification, we run BO with target conductivity on a artificial ground truth experiment whose data generator is an ensemble of graph models trained by our final CNT dataset. In detail, measuring the conductivity of one droplet of a given CNT content $C$ is corresponding to one call of $Y$ by a graph model with input $C$. For the graph model, we use the top 10 graphs selected from the candidate graphs using the method given in Section 3.3. For each graph, we train 20 graph models using randomly resampled training dataset. During one call of $Y$, we first randomly choose one of the trained graph models, and then use it to predict $Y$. hyper-parameters same with our real experiment:

batch_size = 9,

num_cores = 4,

model_type='GP',

initial_design_numdata=20,

acquisition_type ='EI',

acquisition_jitter = 0.5.

We independently run the BO 20 times and give their experimental BO history in Figure S6. In each run, we optimized the conductivity up to step = 20. Each box in the boxplot shows the statistical result from all 20 independent runs. Figure S6(a) shows the target convergence by considering the largest observed conductivity in each step (note that there are 9 values in each step since we use the batch-BO). The red circles show the best conductivities by each BO run. Most BO runs get the maximal (noisy) conductivity before step=13. The improvement is obvious in the first 3 steps. Figure S6(b) shows the convergence of the parameter (characterized by the distance to the best obtained content which has the optimal observed conductivity over all runs and steps). The red circles show the nearest distance by each BO run. The minimal distances are less than 10 for most runs. The target and parameter convergence results suggest that around 12 BO steps are enough for our CNT experiment. We attribute this fast convergence to the relative simplicity of the target landscape.

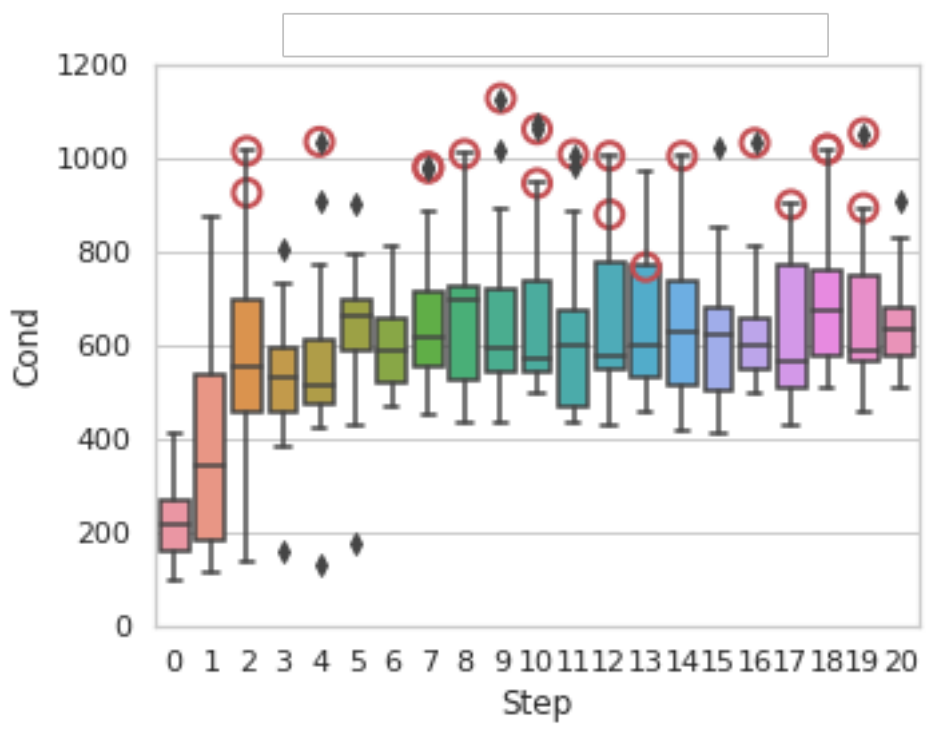

(a)

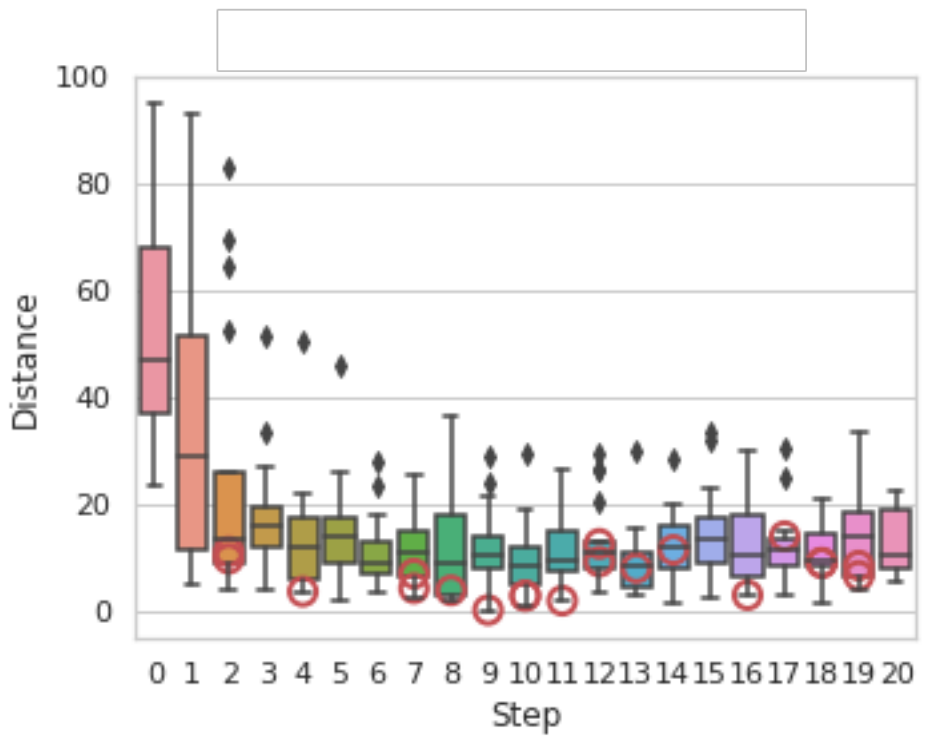

(b)

*Figure S6 BO convergence*

## 5.9 3.5 Related machine learning methods

The machine learning model proposed in this paper is graph-based that handles multiple possible inputs, outputs and intermediate relationships. Our method is not only a regression model (with a fixed graph) but also a model selection strategy considering both accuracy and uncertainty, with the eventual task of determining an optimal graphical model on which to build our regressor.

## Missing data

During the dataset generation step, we complete the missing values by regression, which is related to the general problem of matrix completion[5–8]. However, for our CNT problem, the performance using those matrix completion algorithms (implemented by the 'fancyimpute' package: https://github.com/iskandr/fancyimpute) is similar with linear regression.

## Multi-target regression

From the regression aspect, our graph-based model (GM) is most related to multi-target regression (MTR), also known as multivariate or multi-output regression[9–12]. MTR problems arise in various fields[10]. In contrast to traditional regression tasks where the target variable is single (or multi-targets are treated as separate ones), MTR attempt to exploit the inter-target dependencies between the targets. Two state-of-the-art MTR approaches are Stacked Single Target (SST) and Ensemble of Regressor Chains (ERC)[10]. SST and ERC can be regarded as special cases of GM, where the SST approach for our CNT dataset corresponds to the graph 'C→ARSTY*, CARSTY*→Y', and ERC corresponding to an ensemble of graphs like 'C →A, CA→R, CAR→S, CARS→T, CARST→Y' (with the chain 'ARSTY'). The main differences in our approach are that (1) we focus on one target (Y) and other targets are potential side-information to improve the regression performance, and thus not strictly a multi-target problem; (2) we take the data uncertainty and incompleteness into account, which comes from the noise and limited cost in measurements; (3) our method is not only a regression model (with fixed graph) but also a model/feature selection strategy, with the additional task of determining an optimal graphical model on which to build our regressor. The last point connects our method to the study of inferring dependencies on a graph, namely probabilistic graphical models.

## Uncertainty estimation

During the graph-based model selection, we use an additional KL-score related to the uncertainty estimation, which is a widely studied topic in machine learning, especially in Bayesian learning[13,14]. However, we do not require a very accurate characterization of the distribution of uncertainty. We only need an approximate measure of the noise magnitude to help us perform model selection. As we have to loop through many graph models, efficiency becomes extremely important.

## Probabilistic graphical models (PGM)

PGM[15] combines the rigor of a probabilistic approach with the intuitive representation of relationships given by graphs. They are commonly used in probability theory, statistics, and machine learning. Two branches of graphical representations of distributions are widely used, namely, Bayesian networks and Markov random fields. However, PGM's goal is to infer conditional dependence/independence. In our case, we want to maximize predictive accuracy, i.e., we can ignore conditional dependence if they do not contribute to a better predictive model. This induces a subtle but important distinction, and a mathematical study of the precise distinctions between probabilistic graphical models and graphical regression models (such as the one introduced here) is a subject of future work.

**4. ML validation by high-fidelity experiments**

To further understand the relation between the sample composition and electrical conductivity, four samples are selected for detailed experimental analysis using UV-VIS-MIR spectroscopy (Shimadzu UV-3600), Raman Spectroscopy (532nm laser excitation) and scanning electron microscopy (FEI Helios 600). The samples include two high performance composites: $L_{60}P_{40}$ (60% long wall CNT and 40 % P3HT), $L_{50}D_{10}P_{40}$ (50% long wall CNT, 10 % double wall CNT and 40 % P3HT), a moderate performance composite $L_{10}P_{90}$ (90 % P3HT and 10 % long single wall CNT), and $S_{60}P_{40}$, a low performance composite (with 40 % P3HT and 60 % short single wall CNT). The full absorption spectra of the pristine and doped films are shown in figure S8. The spectrum of P3HT is also included for reference.

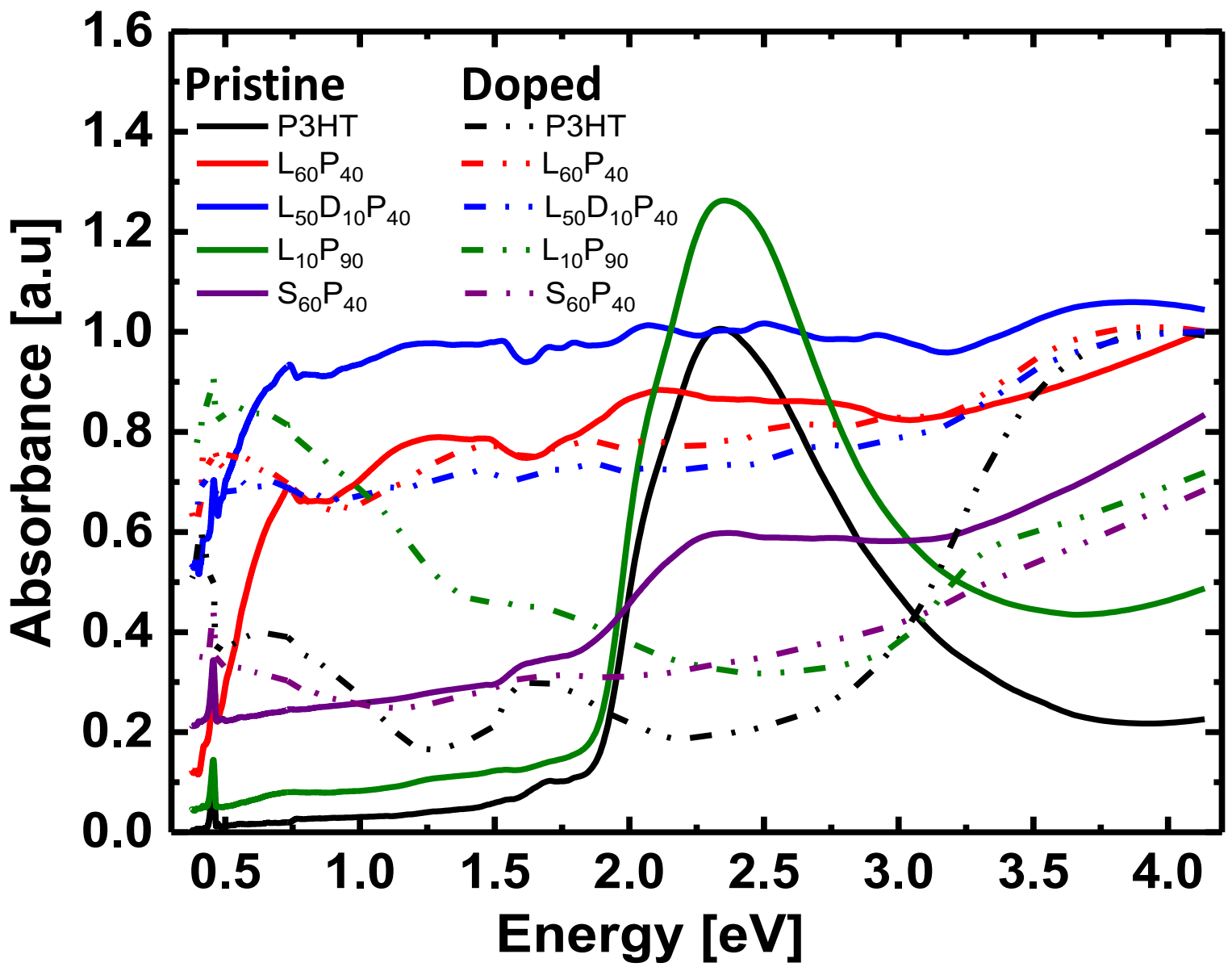

*Figure S7 UV-VIS-MIR spectra of pristine and doped films. Solid lines are for pristine films and dotted lines are for Fecl₃ doped films*

The Raman spectra of pristine CNT films are shown in figure S9 where the contributions due to G-band and D band can be seen clearly. The Raman spectra of P3HT/CNT hybrid films and discussion are included in the main manuscript.

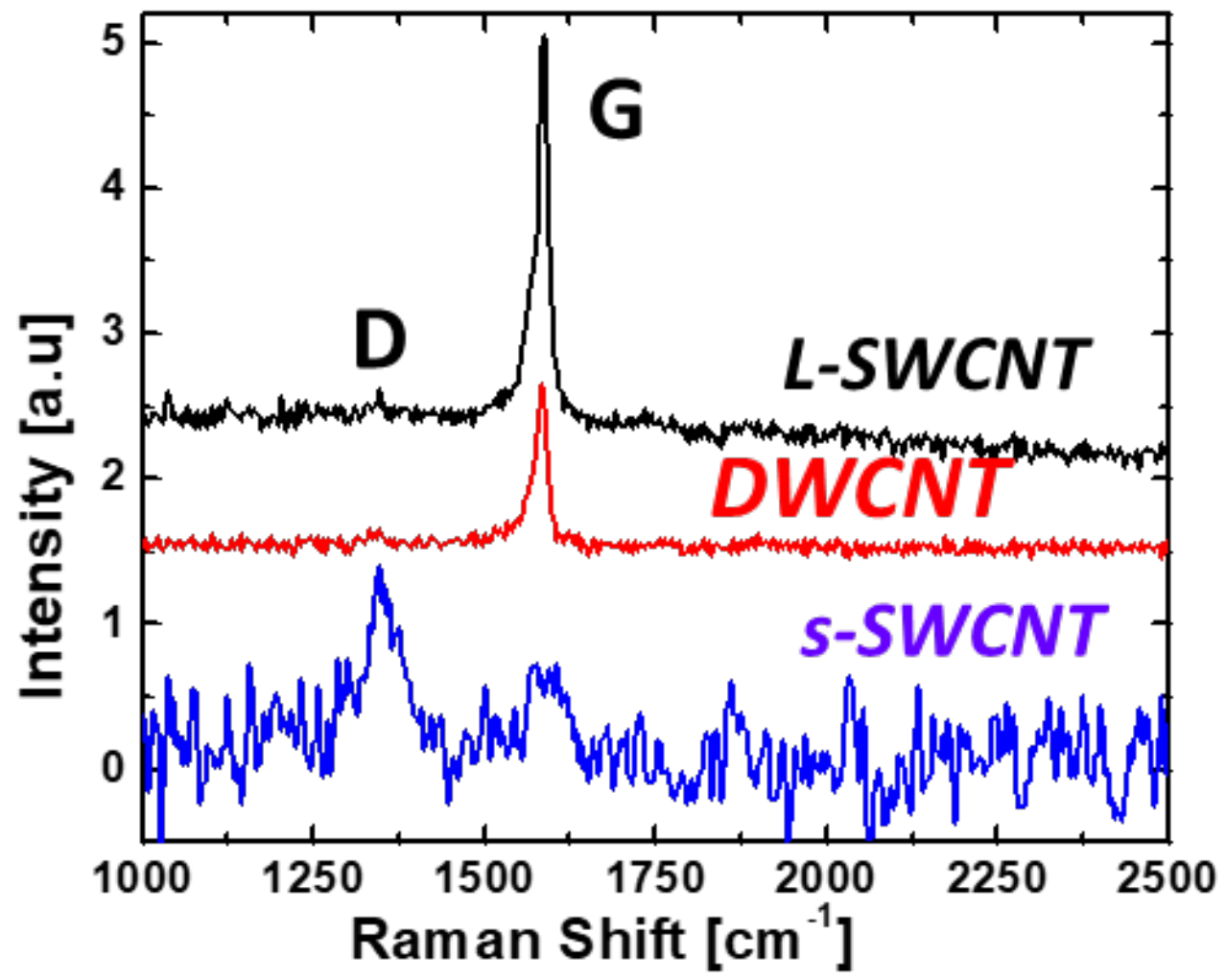

*Figure S8 Raman spectra of pristine CNT films*

The SEM picture of the doped films are shown in figure S10, where the film morphology differences between the different composites can be seen clearly.

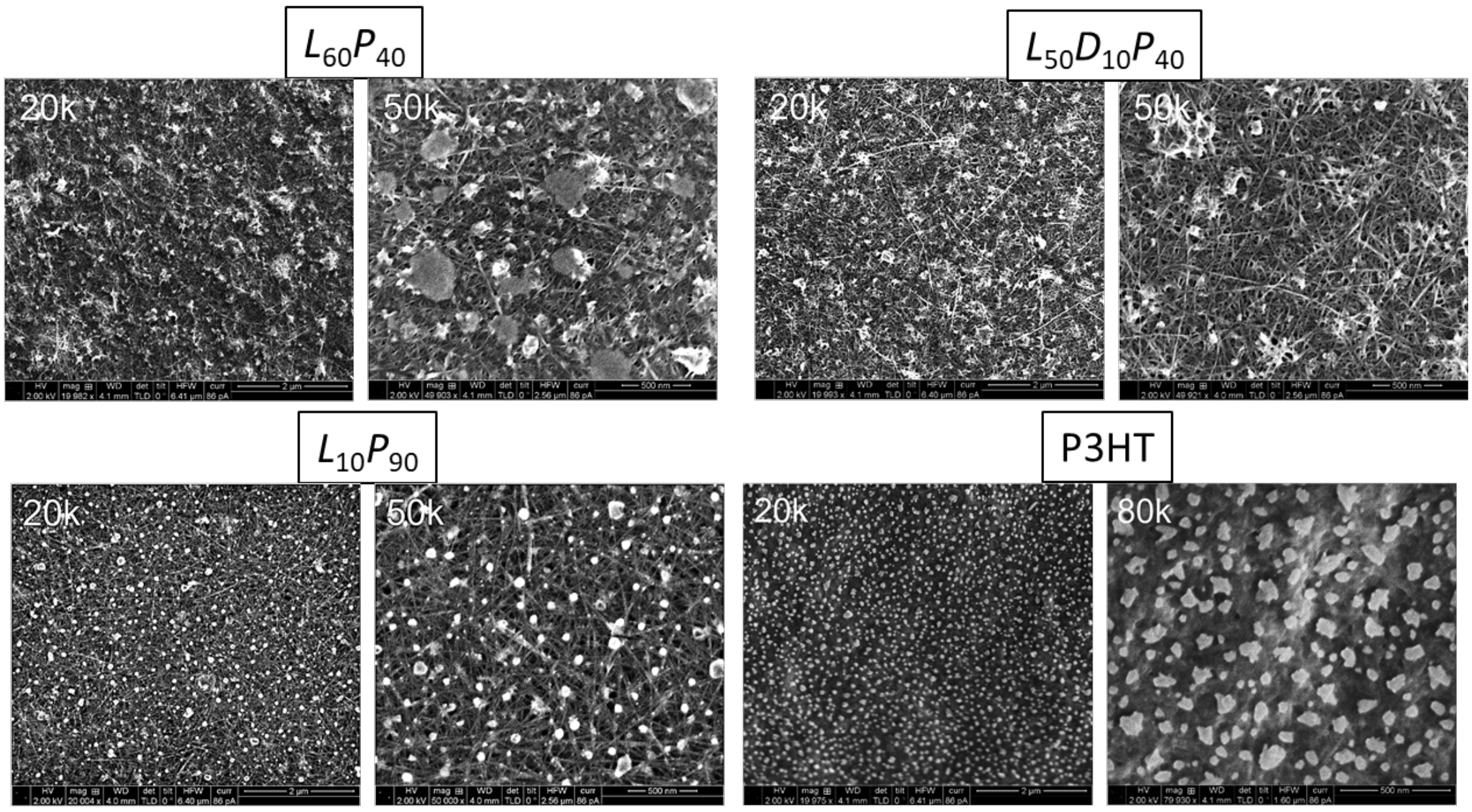

*Figure S9 Contrast enhanced SEM picture of doped films*